\documentclass[12pt,preprint]{aastex}

\usepackage{natbib}
\shorttitle{RR Lyraes in the NSVS}
\shortauthors{Kinemuchi}

\begin{document}

\title{Analysis of RR Lyrae Stars in the Northern Sky Variability Survey}

\author{K. Kinemuchi\altaffilmark{1} and H.A. Smith}
\affil{Department of Physics \& Astronomy, Michigan State University, 
East Lansing, MI 48824}

\author{P. R. Wozniak}
\affil{Los Alamos National Laboratory, Los Alamos, NM 87545}

\author{T. A. McKay}
\affil{Department of Physics, 2477 Randall Laboratory, University of Michigan, Ann Arbor, MI 48109}

\and

\author{ROTSE Collaboration}

\altaffiltext{1}{present address: University of Wyoming, Department of Physics \& Astronomy, Dept. 3905, Laramie, WY 82071} 

\begin{abstract}

We use data from the Northern Sky Variability Survey (NSVS), obtained
from the first generation Robotic Optical Transient Search Experiment
(ROTSE-I), to identify and study RR Lyrae variable stars in the solar
neighborhood.  We initially identified 1197 RRab (RR0)
candidate stars brighter than the ROTSE median magnitude $V = 14$.
Periods, amplitudes, and mean $V$ magnitudes are determined for a
subset of 1188 RRab stars with well defined light curves.
Metallicities are determined for 589 stars by the Fourier parameter
method and by the relationship between period, amplitude, and [Fe/H].
We comment upon the difficulties of clearly classifying RRc (RR1)
variables in the NSVS dataset.  Distances to the RRab stars are
calculated using an adopted luminosity-metallicity relation with
corrections for interstellar extinction.

The 589 RRab stars in our final sample are used to study the
properties of the RRab population within 5 kpc of the Sun.  The Bailey
diagram of period versus amplitude shows that the largest component of
this sample belongs to Oosterhoff type I.  Metal-rich
($[Fe/H] > -1$) RRab stars appear to be associated with the Galactic
disk.  Our metal-rich RRab sample may include a thin disk as well as a
thick disk population, although the uncertainties are too large to 
establish this.  There is some evidence among the metal-rich RRab
stars for a decline in scale height with increasing [Fe/H], as was
found by \citet{Layden:1995}.  The distribution of RRab stars
with $-1 < [Fe/H] < -1.25$ indicates that within this metallicity range
the RRab stars are a mixture of stars belonging to halo and disk
populations. 

\end{abstract}


\section{Introduction}

RR Lyrae variable stars are versatile objects for astronomical
research.  Not only are they excellent distance indicators, but they
are useful as probes for understanding Galactic evolution and
structure.  RR Lyrae variables have the virtues of being both luminous
and ubiquitous in the Galaxy, tracing the old stellar populations of
the bulge, disk, and halo components.  In this paper, we present
results from a new survey of Bailey ab-type RR Lyrae stars (RRab or RR0) in
the solar neighborhood based upon data in the Northern Sky Variability
Survey \citep{Wozniak:2004a}, hereafter NSVS.  Bailey type c (RRc or RR1)
variables in the NSVS will be discussed in a future paper.

To get a clearer picture of the Galaxy, we need as complete a database
as possible of both field and globular cluster RR Lyrae stars.  The
General Catalog of Variable Stars (GCVS) \citep{Kholopov:1996}
summarizes much previous work on the discovery of RR Lyrae stars in
the Galactic field, but is significantly incomplete, especially for RR
Lyrae stars of low amplitude \citep{Akerlof:2000}.  More recent CCD
surveys of RR Lyrae stars in the Galactic bulge (OGLE: \citet{OGLE},
MACHO: \citet{MACHO2}) and halo (SDSS: \citet{SDSS}, QUEST:
\citet{Vivas:2004}) have extended the coverage of the GCVS, but do not
emphasize the discovery of brighter RR Lyrae stars in the solar neighborhood.  

The Northern Sky Variability Survey \citep{Wozniak:2004a} is based on
photometric observations obtained with the first generation telescope
of the Robotic Optical Transient Search Experiment (ROTSE-I) (see
section 2).  Within the NSVS, RR Lyrae stars as faint as 15th
magnitude in V can be detected, extending to a distance of about 7-9
kpc from the Sun \citep{Akerlof:2000}.  This region includes part of
the inner halo and thick disk components of the Galaxy. Hence, this
survey is an excellent complement to the bulge and outer halo surveys,
bridging the portions of the Galaxy that they cover.  The NSVS data 
complement those of the QUEST survey \citep{Vivas:2004}, which includes
RR Lyrae stars from 4 to 60 from the sun, but principally samples RR Lyrae
stars more distant than those in the NSVS.  Another survey that
will complement the NSVS is the All Sky Automated Survey (ASAS)
\citep{Pojmanski:1997}.  The ASAS database surveys the southern sky up
to $\delta = +28$ and has a magnitude limit of $V = 14$.  

In this paper, we will limit ourselves to a consideration of those
NSVS RRab stars that are brighter than 14th magnitude in the ROTSE-I
system.  These stars have light curves sufficiently well defined that
the periods are secure, and the classification of Bailey type is
clear.  Within certain limits, which we shall describe later, we
expect this survey to provide a more complete and unbiased catalog of
the RRab stars in the solar neighborhood than is provided by the GCVS.
Because of this magnitude restriction, our paper mainly concerns RRab
stars to a distance of 5 kpc from the Sun.

Section 2 of the paper briefly describes the ROTSE-I telescope and
the NSVS database.  In section 3, we discuss the selection
criteria used to identify RRab stars, the determination of periods,
the photometric calibration, and methods for determining photometric
metallicities.  In section 4, the properties of the RRab sample and
its distribution in space are discussed. Results are summarized in
section 5.

\section{NSVS Database}

The NSVS database\footnote{See http://skydot.lanl.gov} is constructed
from observations taken from the first generation Robotic Optical
Transient Experiment (ROTSE-I).  The ROTSE-I telescope, located in Los
Alamos, New Mexico, was originally designed to find the optical
counterparts to gamma ray bursters.  While waiting for a trigger from
a gamma ray burster, the telescope was set to a patrol mode, which
scanned the sky.  The NSVS is based on a year's worth of observations,
beginning in April of 1999 and ending in March 2000.

The telescope, now retired, was comprised of 4 smaller telescopes
mounted on a rapidly slewing platform.  Each telescope of the $2
\times 2$ array of telescopes was equipped with a $2048 \times 2048$
CCD.  The main optical element of each telescope was a Canon FD 200mm
f/1.8 telephoto lens \citep{Kehoe:2001}.  The plate scale was
$14.4\arcsec$ per pixel, which corresponded to a $16.4\degr \times
16.4\degr$ field of view.  Each individual telescope had a field of
view of $8.2\degr \times 8.2\degr$.  Thus, the sky was divided into
206 fields (see Figure 1 of Akerlof et al. 2000b).  Each night the
visible fields were observed in pairs of 80 second exposures, while in
patrol mode. In some cases, multiple pairs of observations were taken
for some fields in a single night.

The ROTSE-I telescope did not have a standard filter set installed.
The instrumental photometric bandpass therefore does not exactly
correspond to those of any of the Johnson-Cousins filters, although
the peak sensitivity of the system is similar to that of the Cousins R
band.  The calibration of instrumental magnitudes to Johnson V will be
discussed in section 3.4.  The basic data reductions, i.e. bias/dark
subtraction and flat fielding, source detection, centroiding and
preliminary calibrations of positions and magnitudes were performed
using the legacy ROTSE-I software \citep{Akerlof:2000}.  The ROTSE-I
pipeline performs source extraction and flux measurements using the
SExtractor package \citep{Bertin:96}.  The astrometric and photometric
calibration of each image is based on approximately 1500 field stars
from the Tycho catalog.  NSVS magnitudes are unfiltered ROTSE-I
magnitudes converted to a V-band scale of a mean Tycho star.  The
details of the light curve formation, the photometric corrections, and
data quality indicators are given in \citet{Wozniak:2004a}.  The NSVS
data processing flags capture a number of potential problems with the
measurements and helped to identify observations suitable for our
analysis.  In our work, we have converted the modified Julian dates
listed in the NSVS to heliocentric Julian dates.

\section{RR Lyrae Data}

The data used for this work were obtained from a version of the NSVS
database at the University of Michigan.  At the University of
Michigan, a cross correlation of the variable star candidates with the
2MASS database had been completed.  This gave us additional
information for the selection of RR Lyrae candidates.  The results
presented in this paper is based on the data obtained from this
version of the NSVS database.

\subsection{Selection Criteria}

To identify whether a star in the NSVS database was a variable star, a
variability index was used.  The index is based on previous work
outlined in \citet{Akerlof:2000}.  The variability index itself is
based on the work of \citet{WS:93} and \citet{Stetson:87,
  Stetson:96}.  The main variability index used in this work is
Stetson's L index, which is constructed from Stetson's J and K
indices, which are defined below.  This particular method of finding
variable stars was chosen because the L variability index is optimized
for pairs of observations.  This method also allows us to give a lower
weight to spuriously bright or faint measurements and to avoid
misidentification of non-variable stars.  This index helps exclude
transient events such as flare stars, which, though interesting in
their own right, are not the focus of this paper.

Three separate calculations must be completed to arrive at the final L
index. The first calculation is of the J index.  The J index is
constructed from the observations and their uncertainties.  The {\it
  sgn} function provides the positive or negative sign of the quantity
to which it is applied.
\begin{equation}
J = \frac{\sum_{k=1}^{n} w_{k} sgn(P_{k}) \sqrt{\left|P_{k}\right|}}{\sum_{k=1}^
{n} w_{k}}
\end{equation}

\noindent The weighting factor, $w_{k}$, is set to either 1 or 0.5
depending on whether a pair of observations or only one of the
observations of the pair was usable (as determined from the data
reduction flags).  The expression $P_{k}$ is a product of residuals,
or ``relative'' errors, of two observations, i and j.  
\begin{equation}
P_{k}= \delta_{i_{\scriptstyle k}} \delta_{j_{\scriptstyle k}}
\end{equation}

The individual residuals can be defined for one bandpass or two.  In
the case of the ROTSE-I observations, the bandpass is the ROTSE V
magnitude.  The weighting factor in the residual is derived from the
number of total observations, {\it n}.  For the calculation of the J
variability index, only the pairs of good quality observations were
used, thus the weighting factor was set to 1.
\begin{equation} \label{residual_eq}
\delta_{i} = \frac{v_{i} - \bar{v}}{\sigma_{v_{\scriptstyle i}}} \left(\sqrt{\frac{n}{n-1}}\right)
\end{equation}

The mean magnitude, $\bar{v}$, is calculated with a special weighting
factor.  This weighting factor helps reduce the influence of 
any outliers in a set of measurements \citep{Stetson:87, Stetson:96}. 
The benefit from using this particular weighting factor is that it
changes continuously with the data.  This feature prevents the loss of
information from an outlying observation.  However, any of those
outliers are weighted appropriately and contribute a small amount
toward the mean magnitude.
\begin{equation} 
w = \left[ 1 + \left( \frac{\left|\delta \right|}{a} \right)^b
  \right]^{-1}
\end{equation}

The parameters {\it a} and {\it b} are set to equal 2.0, and $\delta$ is the
residual error as defined in Equation \ref{residual_eq}.  Those
magnitude measurements with large uncertainties or observations that
are unusually bright (or faint) are given a smaller weight in the
calculation of the mean.  Use of a weighted mean magnitude defined in
this fashion gives our method greater sensitivity in the detection of
variable stars, but excludes objects such as flare stars and detached
eclipsing binaries \citep{Akerlof:2000, Stetson:96}.  This weighting
factor is iterated a minimum of five times in order for the mean
magnitude to stabilize in value. 

The second variability index that was calculated is the kurtosis
index, or Stetson K index.  This index is also constructed from the
residuals of Equation \ref{residual_eq} and {\it N} is the number of
observations without regard to the pairs of observations.
\begin{equation}
K = \frac{\frac{1}{N} \sum_{i=1}^{N} \left|\delta_{i}\right|}{\sqrt{\frac{1}{N} 
\sum_{i=1}^{N} \delta_{i}^{2}}}
\end{equation}

The third calculation is the main L variability index, which is
constructed from the J and K indices \citep{Stetson:96}.
\begin{equation}
L = \left( \frac{JK}{0.798} \right)\left(\frac{\sum w}{w_{all}}\right)
\end{equation}

\noindent The weighting factor of this robust variability index ($\sum
w / w_{all}$) normalizes the weights and accounts for any
non-observation of the star from any image.  A non-observation could
occur from the star falling near the edge of the chip or in a cosmetic
blemish on the chip.  However, in the case of the ROTSE-I survey,
there is a reasonable amount of overlap between fields, so edge drop
offs are not a problem.  Multiple observations of the same star can
occur if there are tiny differences in the astrometry.  This
multiplicity was first documented in \citet{Wils:2001} for the
variables stars reported in \citet{Akerlof:2000}.  The flag system
(see \citet{Wozniak:2004a}) accounted for these two CCD detection
problems, so we were able to avoid observations where this may have
occurred.  Thus only the ``good'' observations are passed to the
calculation of the L index.  With this consideration, the weighting
factor was set to 1.  The factor in the denominator, $w_{all}$, was
also set to 1 since it represents the total weight of the star if it
was observed in all frames \citep{Stetson:96}.  Therefore, the final
index used to identify the variable stars is simplified to
\begin{equation}
L = \frac{JK}{0.798}
\end{equation}

\noindent The value of 0.798 in the denominator comes from the
statistics of the K index \citep{Stetson:96}.

A cutoff value had to be chosen for the L index, above which the star
could be considered a variable star candidate.  This cutoff value is
dependent on magnitude and position.  Thus the L index cutoff value
was determined for each ROTSE-I field.  Trends across the sky were
investigated as the index is dependent on the number of observations
taken for each star.  The northern fields were observed more than the
southern fields due to the location of the telescope, and therefore
values for the L index were more robust for the northern fields.
Fewer observations were also made for stars found near the Galactic
plane due to crowding issues.  These factors contribute to the calculation
of the variability index value and to the selection of the appropriate
cutoff value.

\begin{figure}
\figurenum{1}
\includegraphics{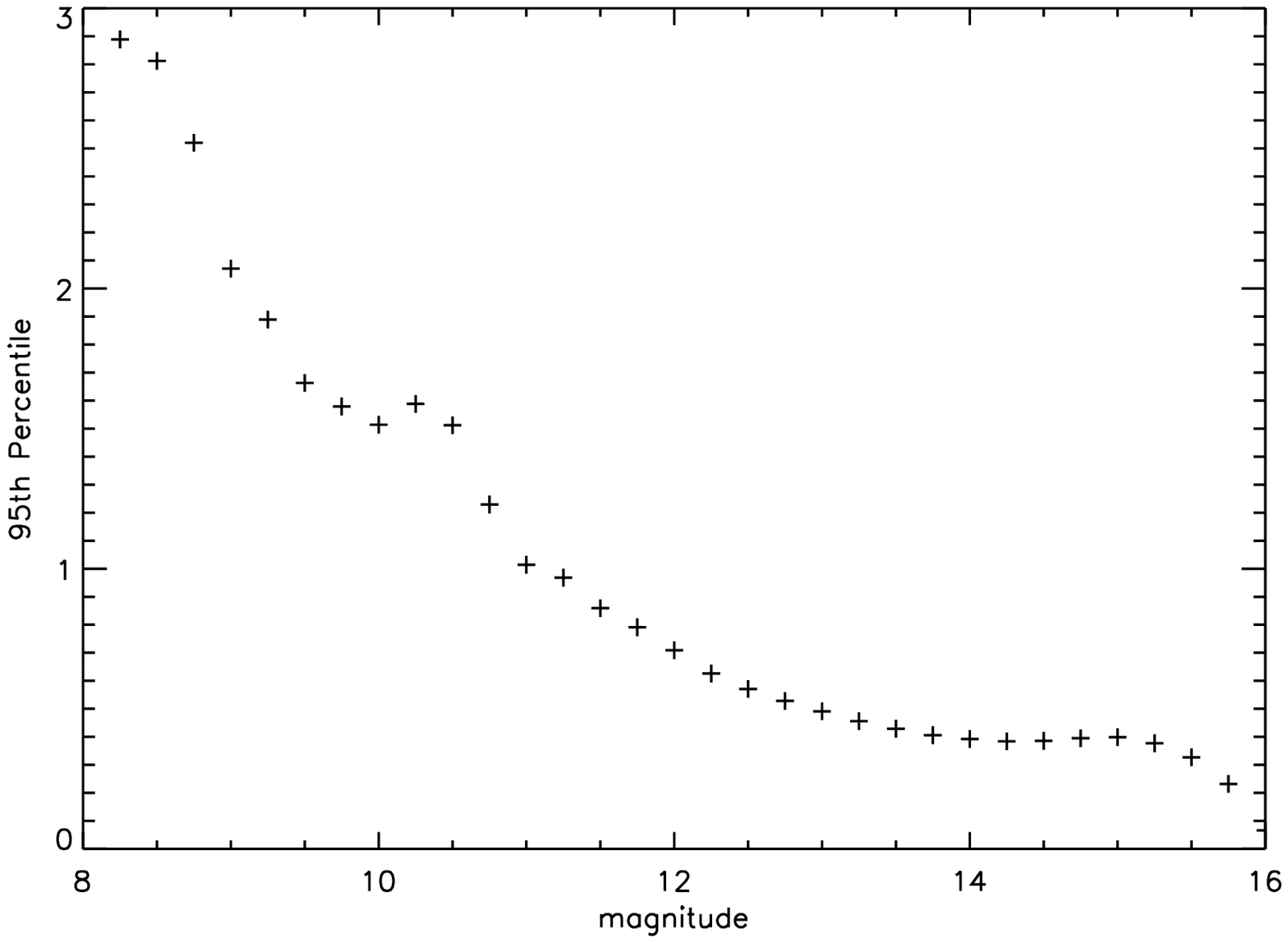}
\caption{The variability index cutoff has biases with magnitude and
  position as described in section 3.1.  To determine the cutoff
  values, the index was binned with respect to magnitude.  The
  abscissa is the 95th percentile value of the variability index per
  magnitude bin.}
\label{polar95}
\end{figure}

\begin{figure}
\figurenum{2}
\includegraphics{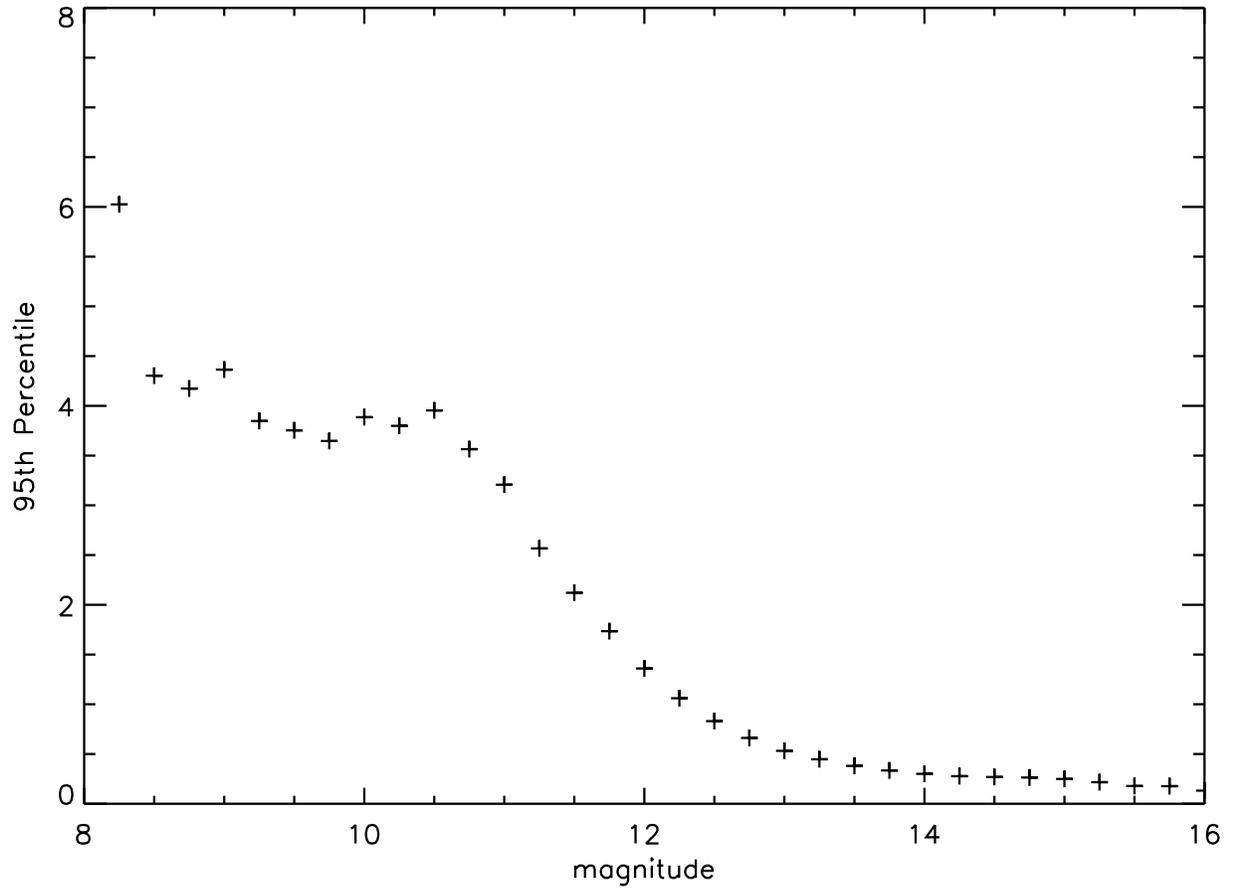}
\caption{The same variability index plot as Figure \ref{polar95}, but
  for a test field near the Galactic plane.}
\label{planar95}
\end{figure}

In addition to these positional effects to the determination of the L
index cutoff value, the influence of magnitude was also investigated.
The magnitude effect was analyzed in two test fields, one in the
northern portion of the survey and the other closer to the Galactic
plane.  The stars in each test field were binned by magnitude.  The
brighter stars, about 10th magnitude or brighter, were found to have
mostly overestimated photometric uncertainties, whereas the fainter stars (15th
magnitude or fainter) had underestimated uncertainties.  In each
magnitude bin, the 95th percentile value of the stars' L indices was
determined to be the L index cutoff.  Next a polynomial fit was
performed on each cutoff value per magnitude bin to see the
behavior of the L index with magnitude.  All stars above the cutoff
value per magnitude bin were identified as variable stars.  In Figure
\ref{polar95}, we see how the variability index behaved in a test field
near the Galactic pole, and in Figure \ref{planar95}, for a test field
near the Galactic plane.  In both plots, we have binned the
variability index by magnitude. The abscissa in each plot is the 95th
percentile value of the variability index within each magnitude bin.

Now with the behavior of the variability index within the NSVS survey
understood, the variable stars can be identified.  An initial
selection of RRab stars was performed on the NSVS database at the
University of Michigan.  These stars, as well as other types of variables,
were cross correlated with the 2MASS catalog.  The 2MASS observations
were taken simultaneously in the J, H, and K bands, and thus, the
colors $(J-H)$ and $(H-K)$ are essentially at some (random) phase of
the variable star's light cycle.  With the addition of the 2MASS
colors, a fine tuned set of parameters was chosen to select a clean
sample of RRab stars.  These parameters included period, amplitude,
2MASS color ($J-H$ and $H-K$), and a magnitude ratio criterion.  The
selection criteria are listed in Table \ref{selection_rrab}.  The
magnitude ratio is constructed from the ROTSE-I median, maximum, and minimum
magnitudes of the variable star candidate.  This parameter was used to
identify whether the variable star spent most of its time above or
below the median magnitude, allowing identification of an eclipsing
binary or a pulsating variable star. 

\begin{equation}
\qquad\mbox{magnitude ratio} = \frac{\qquad\mbox{max magnitude}\qquad - \qquad\mbox{median
magnitude}\qquad}{\qquad\mbox{max magnitude}\qquad - \qquad\mbox{min
magnitude}\qquad}
\end{equation}

The sample of NSVS RRab candidate stars originally contained 2196
stars.  Due to the way the stars were originally identified in the
ROTSE-I database (by position), duplicate entries existed for many of
the variables.  This synonym problem still exists in the NSVS database
and should be taken into account when analyzing large samples of data
from this survey. The issue of multiplicity is discussed in the public
release paper \citep{Wozniak:2004a}.  For the purpose of this work, we
did not combine the multiple entries from the RRab candidates, but
rather chose the entry with the most observations.  Combining the
multiple entries often increased the observational scatter, and for
the phased light curves, caused more difficulty in analyzing the light
curve shape.  A total of 228 duplicate entries were found for this
sample of RRab candidate stars.

\subsection{Period Searching}

Several period searching algorithms were tested for the RRab
candidates, which include the Lomb-Scargle periodogram \citep{Lomb:76,
  Scargle:82} and the Analysis of Variance method \citep{Czerny:89}.
The ideal period searching algorithm had to be fast and mostly
non-interactive for such a large dataset.  For these reasons,
some of the algorithms, such as Phase Dispersion Minimization
(PDM) \citep{Stellingwerf:78} had to be rejected as the main period
searching algorithm.  We decided to use the Supersmoother routine
\citep{Reimann:94} for our RRab candidates.  This routine was also
used by the MACHO group \citep{Alcock:95} for their variable star
catalogs.  

The Supersmoother routine quickly calculated the periods for our
sample of RRab candidates.  This routine is essentially a smoothing
routine where the observations are fit as a function of phase to a
range of frequencies \citep{Reimann:94}.  The smoothing algorithms use
a running mean or running linear regression on the data.  The
advantage to this routine is that it is self-adjusting and the light
curve shape does not need to be known a priori.  \citet{Reimann:94}
conducted extensive comparison tests for Supersmoother against other
period searching algorithms, and its advantages and disadvantages are
described in detail in his thesis.

As in many variable star surveys, aliasing problems had to be dealt
with in determining the correct period for the NSVS RRab stars.  One
of the features of the Supersmoother routine was the ``cycles'' number
which provided a better indication to the correct period of the
variable star.  The cycle number is the number of times the
Supersmoother curve crosses the mean value, divided by 2 (K. Cook,
private communication).  For a pulsating variable star, the cycle
number is usually set to 1.  Higher values of the cycle number for a
pulsator were often identified as a period alias solution.  However,
there is some confusion with W Ursa Majoris (W UMa) stars or other
eclipsing binary stars where the two minima are of equal depth.  For
these two types of variables, the cycle number should be 2.  More of
this classification problem is discussed in regards to the c-type RR
Lyrae stars in Section 4.6. 

The process in finding the period solution for our RRab sample, we
mainly relied on the results from the Supersmoother program.  The best
period was usually found within the first 15 guesses generated by
Supersmoother.  As a comparison and check, we also had a cubic spline
solution \citep{Akerlof:94, Akerlof:2000} from NSVS database at the
University of Michigan.  In the cases where the period solutions
between the Supersmoother and cubic spline methods were not in
agreement to within $10^{-3}$ days, the IRAF version of PDM was used
to resolve the true period and distinguish possible alias solutions.
If the PDM solution agreed with one of the period solutions from
either method, then that result was assigned as the best period for an
RRab star.  There were six stars for which the period solution could
not be resolved, even with the use of PDM, and they were subsequently
omitted from the analysis.  For four stars, all three period searching
algorithms could not identify the true period, but provided 3
different solutions.  These stars were also removed from the working
sample.  These situations often occurred when the alias periods
corresponded to frequencies near 2 cycles/day or 3 cycles/day.  Table
\ref{unknown_p} lists the stars where PDM could not distinguish
between the Supersmoother and cubic spline solutions. 

\begin{figure}
\figurenum{3}
\includegraphics{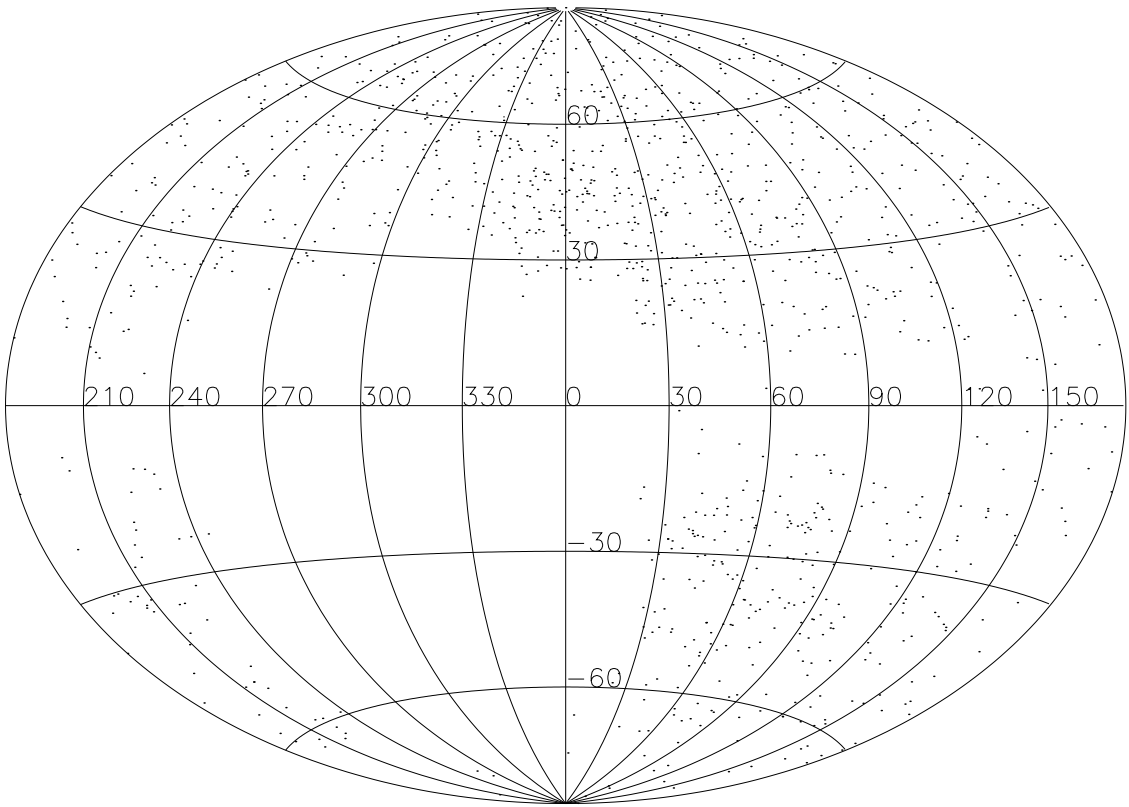}
\caption{The distribution of 1188 RRab stars from the NSVS
  database. The Aitoff plot coordinate system is in Galactic coordinates.}
\label{aitoff}
\end{figure}

\begin{figure}
\figurenum{4}
\includegraphics[angle=-90,scale=0.6]{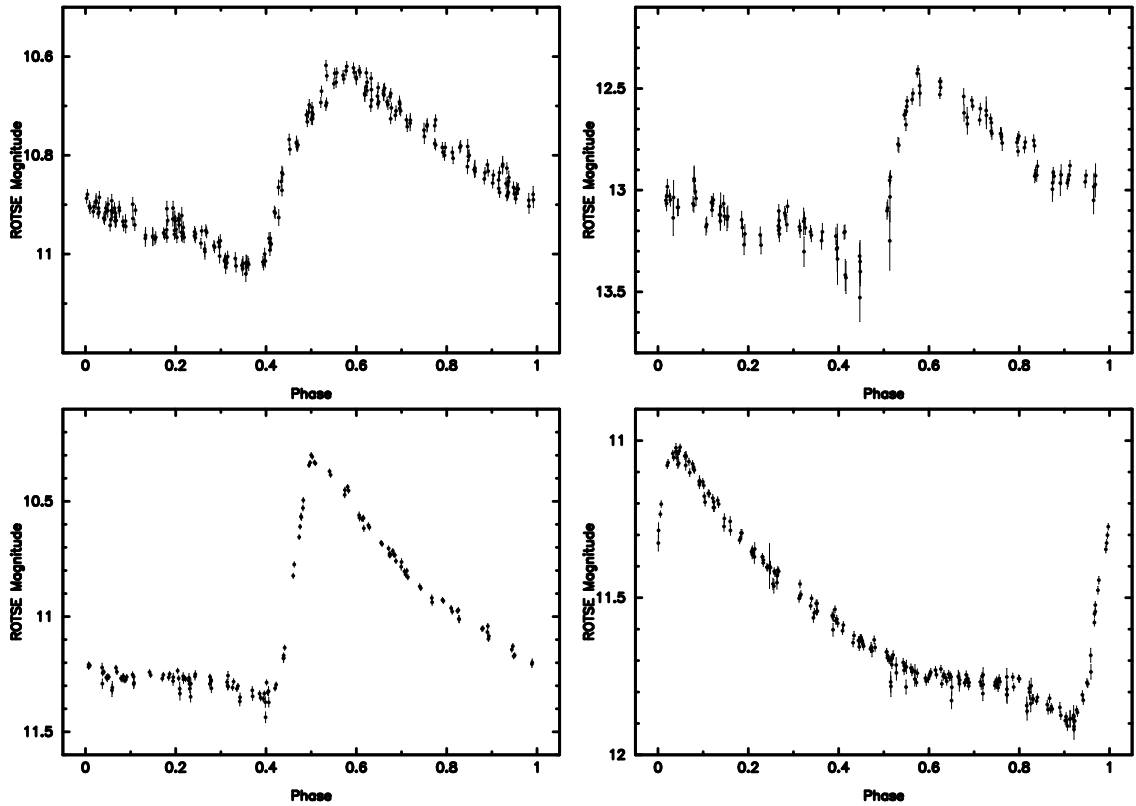}
\caption{Phased light curves of well observed RRab stars.  Note that
  the fainter stars have slightly larger photometric uncertainties.
  The stars' NSVS id numbers, going clockwise and starting from the upper left,
  are 3619356 (AT And), 10514625, 5149127 (TW Boo), and 10706906 (VX Her).}
\label{ltc_rrab}
\end{figure}

\begin{figure}
\figurenum{5}
\includegraphics[angle=-90,scale=0.6]{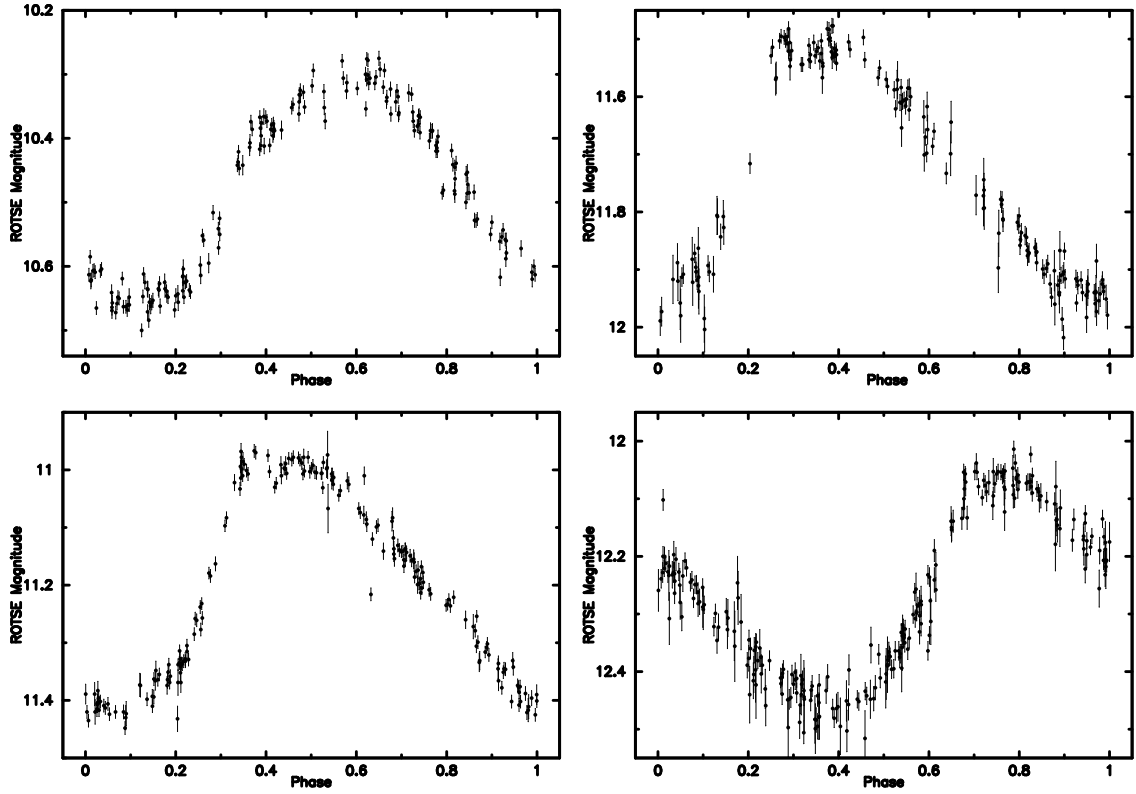}
\caption{Phased light curves of well observed RRc stars.  Note that
  for some of the stars, the bumps associated with shock wave
  phenomena can be clearly seen in these light curves.  The stars'
  NSVS id numbers are, going clockwise and starting from the upper
  left, 6410299 (RU Psc), 7879639 (RV CrB), 6270561 (VZ Peg), and
  2678879 (SX UMa).} 
\label{ltc_rrc}
\end{figure}

With the period information, we visually inspected the phased light
curves of the RRab candidates.  Those stars with an obvious eclipsing
binary star light curve were removed from the sample as well as those
stars with light curves too noisy for analysis.  Removing 771 stars
with these noisy or eclipsing binary light curves, our sample reduced
down to 1197 RRab candidate stars.  With the addition of the period
aliasing problem discussed above, the sample reduced to 1188 stars.
Figure \ref{aitoff} shows the distribution of the 1188 RRab stars in
an Aitoff plot, where the variables are displayed according to their
Galactic coordinates.  The plot also shows the range the ROTSE-I
telescope was able to survey.  The paucity of points near the Galactic
plane is due to the confusion from overcrowding.  Figures
\ref{ltc_rrab} and \ref{ltc_rrc} show examples of well sampled phased
light curves for RRab and RRc stars, respectively. 

\subsection{Amplitude Calibration}

Once all the periods for the RRab candidates were determined,
amplitudes were calculated by two different methods.  A spline curve
fit \citep{numrec:1992} was performed on all the RRab candidates,
yielding amplitudes and intensity-weighted mean magnitudes.  In the
cases where the spline fit was poorly constrained, as in some
instances of unusually poor light curves, a template fitting routine
\citep{Layden:1998, Layden:2000} was also used.  

Amplitudes as measured from the NSVS database are in ROTSE magnitudes
but need to be further calibrated to a standard photometric system.
As mentioned in section 2, the ROTSE-I telescope did not have a
filter.  The peak sensitivity of this instrumental system approximates
to that of a Cousins R value, but it does not actually match
any of the usual photometric filter systems.  Because the ROTSE
observations include a contribution from light redder than that passed
by a Johnson V filter, the amplitudes are smaller than would be
expected if the observations were obtained in Johnson V.  The
amplitude generally decreases when one observes with redder filters
(see discussion in \citet{Smith:95}).  In order to calibrate the
amplitudes, well observed field RRab stars from the GCVS were chosen.
The observed amplitudes were scaled to the published RRab amplitudes in the
literature \citep{Bookmeyer:1977, Kemper:1982}.  Our calibration
relation was determined from 35 RRab and 12 RRc stars. 
\begin{equation}
A_{V} = 1.19(\pm 0.05)A_{NSVS} + 0.07(\pm 0.03)
\end{equation}

The RRab sample was now subjected to the criterion that the RRab
candidate should have at least 40 observations.  The periods, as well
as the amplitudes, for stars with such a low number of observations
could be spurious since the maximum or minimum were often missed.  These stars
also often had gaps in the phase coverage of the light curve, which
excludes them from any later analysis of the light curve shape.
Variable candidates with a mean magnitude fainter than 14th magnitude
were also omitted from our sample.  The quality of the light curves becomes
poorer with increasing magnitude, and although the detection of RRab
stars fainter than 14th magnitude is in many cases possible, the
results are more uncertain than with the 14th magnitude limit.

\begin{figure}
\figurenum{6}
\includegraphics[angle=0,scale=0.8]{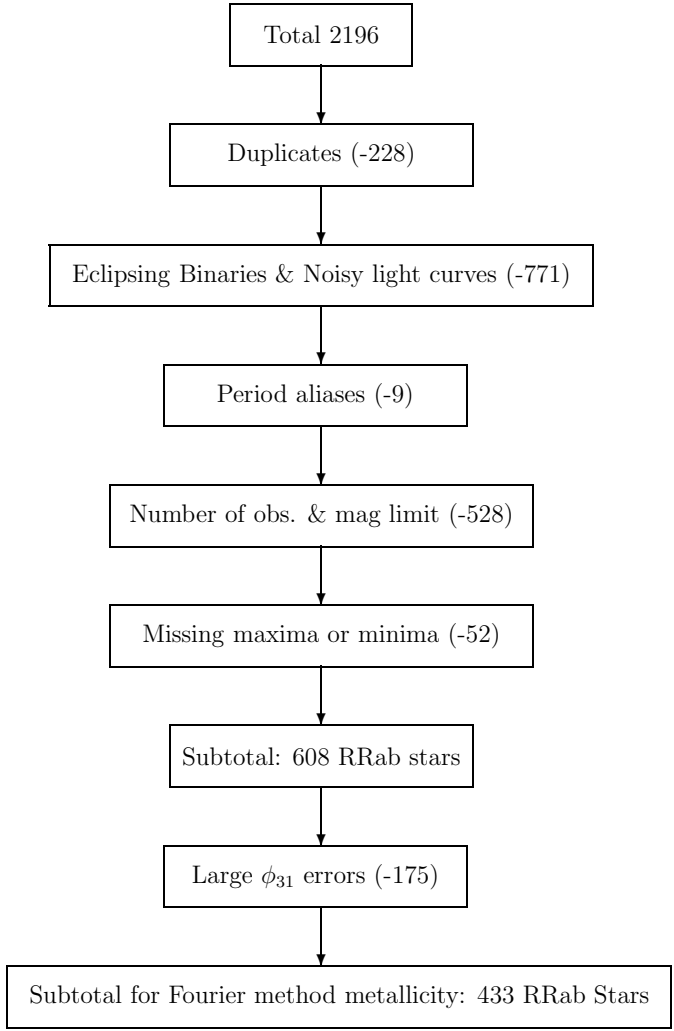}
\caption{Flow chart summarizing the different criteria used to obtain
  a clean sample of NSVS RRab stars.}
\label{flowchart}
\end{figure}

For the RRab candidate sample, about 43\% of the stars were fainter
than 14th magnitude and about 0.7\% were found to have fewer than 40
observations.  Additional observations are needed for those stars with
fewer than 40 epochs.  With these two constraints, the number of RRab
candidates reduces to 660 stars.  However, fifty-two additional stars
were also omitted due to missing observations of either the maximum or
minimum in the phased light curve, although they had more than 40 epochs.  The
total number of RRab stars available for analysis at this point is 608 stars.
A flow chart, presented in Figure \ref{flowchart}, summarizes how we have 
arrived at our RRab sample.

\subsection{Magnitude Calibration}

We used 147 Landolt standard stars \citep{Landolt:1992} observed by the
ROTSE-I telescope to transform the NSVS magnitudes more completely to
the Johnson V system.  A linear regression fit was performed for these
stars, and a transformation equation was derived. 
\begin{equation} \label{vtran_eq}
V_{NSVS} = 0.69(\pm 0.07) + 0.98(\pm 0.01)V_{standard} -
0.66(\pm 0.02)(B-V)_{standard}
\end{equation}

\begin{figure}
\figurenum{7}
\includegraphics[angle=-90,scale=0.6]{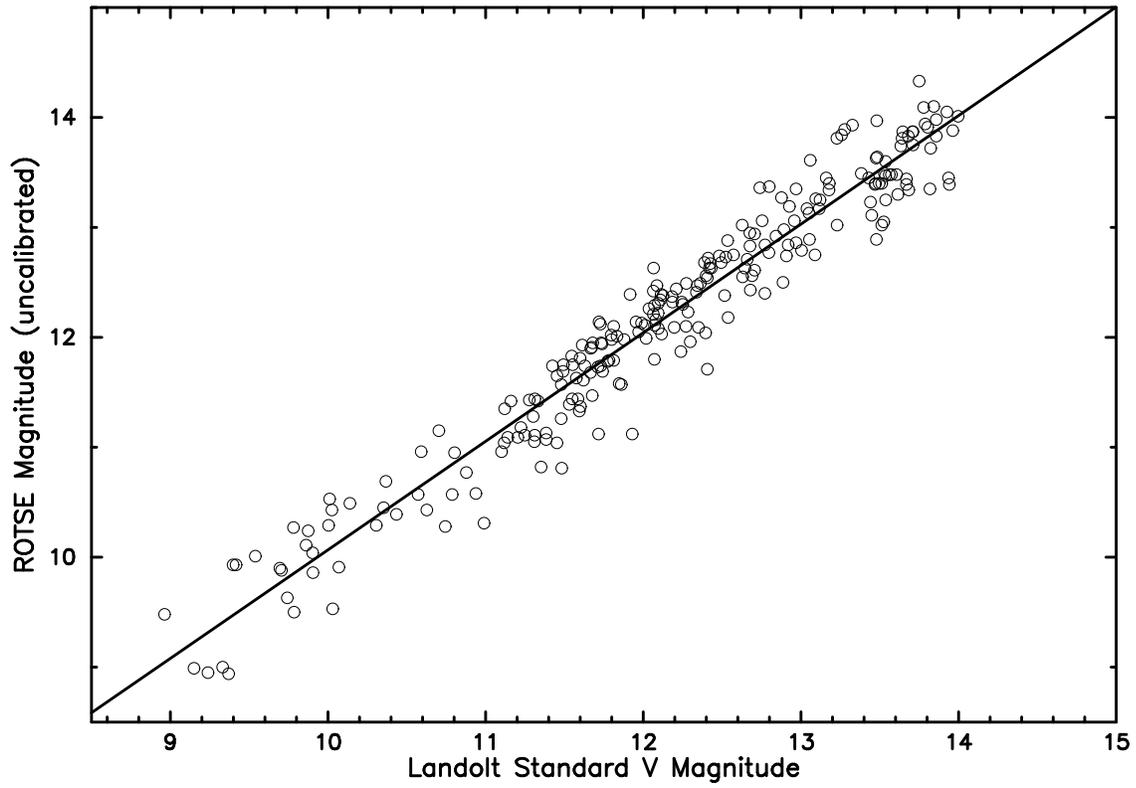}
\caption{The ROTSE magnitudes were calibrated by using 147 Landolt
  standard stars \citep{Landolt:1992}.  This plot shows the calibration
  without accounting for the color term in the transformation.  The
  RMS scatter is 0.28.}
\label{nocolor}
\end{figure}

\begin{figure}
\figurenum{8} 
\includegraphics[angle=-90,scale=0.6]{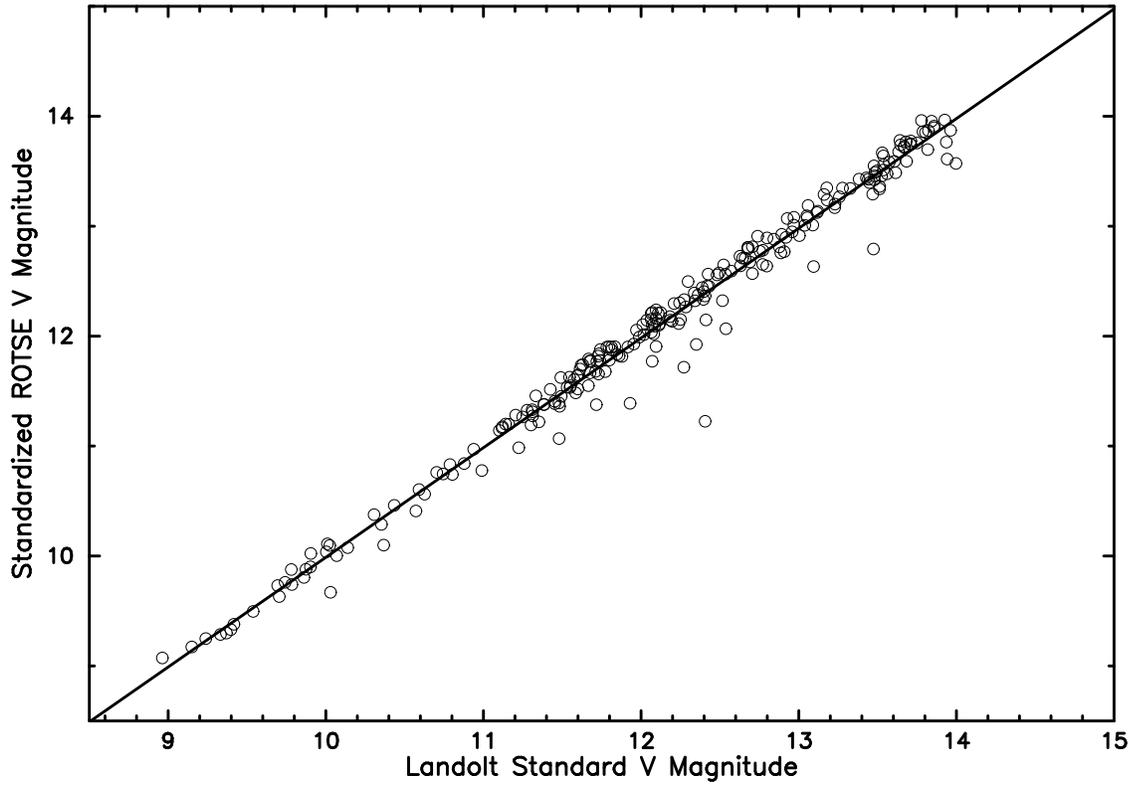}
\caption{The ROTSE magnitude calibration with a color term.  Once the
  color term is accounted for in the transformation equation (Equation
  \ref{vtran_eq}, section 3.4), the calibration of the ROTSE magnitude
  improves.  The RMS scatter here is 0.15.} 
\label{vcolor}
\end{figure}

The typical color of an RR Lyrae is approximately $(B-V) = 0.4$
\citep{Smith:95}, but in determining this transformation relation, the entire
color range of the Landolt stars ($-0.3 < (B-V) < 2.0$) was used.  The
color term was necessary for the calibration, as evident in Figures
\ref{nocolor} and \ref{vcolor}.  Without the color term, the RMS scatter was
0.28, as seen in Figure \ref{nocolor}.  Once the color term was included in the transformation
relation, the V magnitude calibration became tighter with an RMS
scatter of 0.15, as seen in Figure \ref{vcolor}.

\subsection{Fourier Decomposition Parameters and Photometric
  Metallicities}

Traditionally, metallicities of RR Lyrae stars are derived from
spectroscopic observations or from multicolor photometric observations
(e.g. \citet{Preston:59} or \citet{Sturch:66}).  In this large survey
where several hundred of RR Lyrae candidates were detected, it was not
feasible immediately to obtain these observations for all of these stars.  It
would be a great advantage to be able to use the NSVS photometric data
themselves to determine [Fe/H] values.  We employ two methods for
determining the metallicities of RRab variables based upon period
and light curve shape.  The first method depends on the Fourier decomposition
parameters for the light curves.  Using a set of well observed field
RRab stars, \citet{Jurcsik:96} and \citet{Sandage:2004} derived
empirical relations between metallicity, period, and the
$\phi_{31}$ Fourier parameter.  The second method employs the well
established correlation between metallicity and the position of an
RRab star in the Bailey diagram of period versus amplitude
\citep{Preston:59}.  These methods can be applied to all NSVS RRab
variables that have sufficiently well defined light curves.  We note,
however, that both of these empirical relations were
developed for Bailey type ab RR Lyrae stars and cannot be applied to
RRc stars.  Similar photometric metallicity relations for RRc stars
have been proposed by \citet{Morgan:2005}. 

\subsubsection{Jurcsik \& Kovacs Empirical Method}

\citet{Jurcsik:96} used a set of 81 RRab stars in the development their
metallicity relation. The method is a linear relationship of the
period, the Fourier decomposition phase parameter, $\phi_{31}$, and the
metallicity, [Fe/H].  In their sample, Jurcsik \& Kovacs were careful
not to include poor quality light curves or RR Lyrae stars experiencing the
Blazhko effect.  Both of these factors affect the overall shape of the
light curve, and thus produce a poor determination of the Fourier
decomposition phase parameter.  The relation (their equation 4) is
reproduced here.
\begin{equation}
[Fe/H] = -5.038 - 5.394P + 1.345\phi_{31}
\end{equation}

Initially, we calculated the Fourier decomposition parameters with an
8th order fit to each light curve of our sample of 608 RRab stars
using the method described in \citet{Simon:1979} and \citet{Simon:1982}.
Our Fourier parameters were found from a cosine Fourier series fit.
However, Jurcsik \& Kovacs performed a sine Fourier series fit to
their set of RRab light curves for their empirical relation.  Our Fourier
decomposition phase parameters were converted by adding the correct
factors of $\pi$.  The Fourier decomposition phase parameters are
constructed from two phase terms of the fit, $\phi_{ji}=i\phi_{j}
-j\phi_{i}$.  To use the empirical relation of Jurcsik \& Kovacs, we
needed to use the $\phi_{31}$ parameter.  The uncertainty of this
parameter was used as a diagnostic as to whether we could derive a
plausible [Fe/H] value for any particular star.  We found that if the
uncertainty was larger than 0.3, the light curve shape of the star
often had gaps or outliers that affected the fit of the Fourier
series.  After visually inspecting those stars with large $\phi_{31}$
uncertainties, our sample of 608 RRab stars reduced to 433 stars
suitable for application of this method.

The $\phi_{31}$ parameters calculated in this fashion need to be
transformed to the values that would have been obtained had the stars
been observed in the Johnson V bandpass, which was the bandpass used
by \citet{Jurcsik:96}.  For our sample, we used published values of
$\phi_{31}$ for 37 stars from \citet{Sandage:2004} for our calibration
relation.  The corresponding NSVS RRab stars' $\phi_{31}$ is indicated
as $\phi_{31}^{'}$ in the scaling relation:
\begin{equation}
\phi_{31} = 0.855(\pm 0.046)\phi^{'}_{31} + 0.063(\pm 0.105)
\end{equation}

Furthermore, \citet{Sandage:2004} scaled the metallicities obtained
from \citet{Jurcsik:96} relation to the \citet{Zinn:1984} system.
This facilitates a comparison with metallicity values in the
literature for many known RR Lyraes and for globular clusters.
\citet{Sandage:2004} derived the following scaling relation with his
set of field RRab stars, which we adopt:
\begin{equation}
[Fe/H] = 1.05[Fe/H]_{Jurcsik \& Kovacs} - 0.20
\end{equation}

It should be mentioned here that \citet{Jurcsik:96} introduced a
``compatibility test'' to find ``normal'' looking RRab light curves.
The compatibility test consists of finding deviation parameters for
each of the Fourier decomposition parameters.  Updated values for the
deviation parameters are listed in \citet{Kovacs:1998}.  To pass the
compatibility test, the maximum of all the deviation parameters,
$D_{m}$, could not exceed 3.0 \citep{Jurcsik:96, Clement:1999}.  We found
that if we implemented this compatibility test, many of the NSVS RRab
stars did not pass it.  This indicates that most of these stars do not
have ``normal'' light curve shapes.  One likely explanation for the
NSVS RRab stars failing the compatibility test is that the light
curves are constructed from observations not in the V band (on which
this method is based), but rather in an unfiltered ROTSE photometric
system.  The maximum deviation parameter often came from the deviation
parameter for the $\phi_{31}$ Fourier parameter.  In addition, small
gaps or outliers in the light curves could also affect the calculation
of the Fourier decomposition so as to make $D_{m}$ greater than 3.0,
but not by such a large amount as to render the derived metallicities
unusable.  Therefore, we did not employ the $D_{m} = 3.0$ limit as a
guide to select a clean sample of RRab stars. 

\subsubsection{Sandage's Empirical Method}

\citet{Preston:59} found that RRab stars of different [Fe/H] occupy
different locations in the period-amplitude diagram.
\citet{Sandage:2004} later derived an empirical relationship between [Fe/H],
period, and amplitude.  Sandage developed his relation after
investigating the correlation between the \citet{Jurcsik:96} relation
and the period shift of the Oosterhoff dichotomy.  From his set of 50
field RRab stars, Sandage found the Fourier decomposition phase
parameter, $\phi_{31}$, is correlated with the period shift, $\Delta
\log P$, in the same manner as the amplitude or color of the RR Lyrae stars
\citep{SKS:1981}.  With this result, a relation for metallicity that is
dependent linearly with amplitude was derived.
\begin{equation} \label{ampmet}
[Fe/H] = -1.453(0.027)A_{V} -7.990(0.091)\log P - 2.145(0.025)
\end{equation}

To determine the uncertainty of the [Fe/H] values from this method, we
assume that the largest source of error comes from the amplitude
uncertainty, combined with any uncertainty intrinsic to the $\log$
P-$A_{V}$-[Fe/H] relation.  We adopt
an uncertainty of 0.32 dex for [Fe/H] derived through this method.
This estimate of the [Fe/H] uncertainty is based upon a comparison of the
spectroscopically derived [Fe/H] from \citet{Layden:1994} to the [Fe/H]
values derived from Sandage's relation.  We subtracted in quadrature
the uncertainty Layden attributed to his [Fe/H] measurements.

Despite its relatively large uncertainty, an important advantage to
Sandage's empirical relation over the Jurcsik \& Kovacs method is the
use of period and amplitude alone to arrive at an [Fe/H] value.
Obtaining amplitudes from the light curves is less difficult than
deriving accurate $\phi_{31}$ Fourier parameters.  Sandage's method
allows us to use most of the NSVS RRab stars in our sample.

\subsubsection{Best Estimate Metallicity}

\begin{figure}
\figurenum{9}
\includegraphics[scale=0.8]{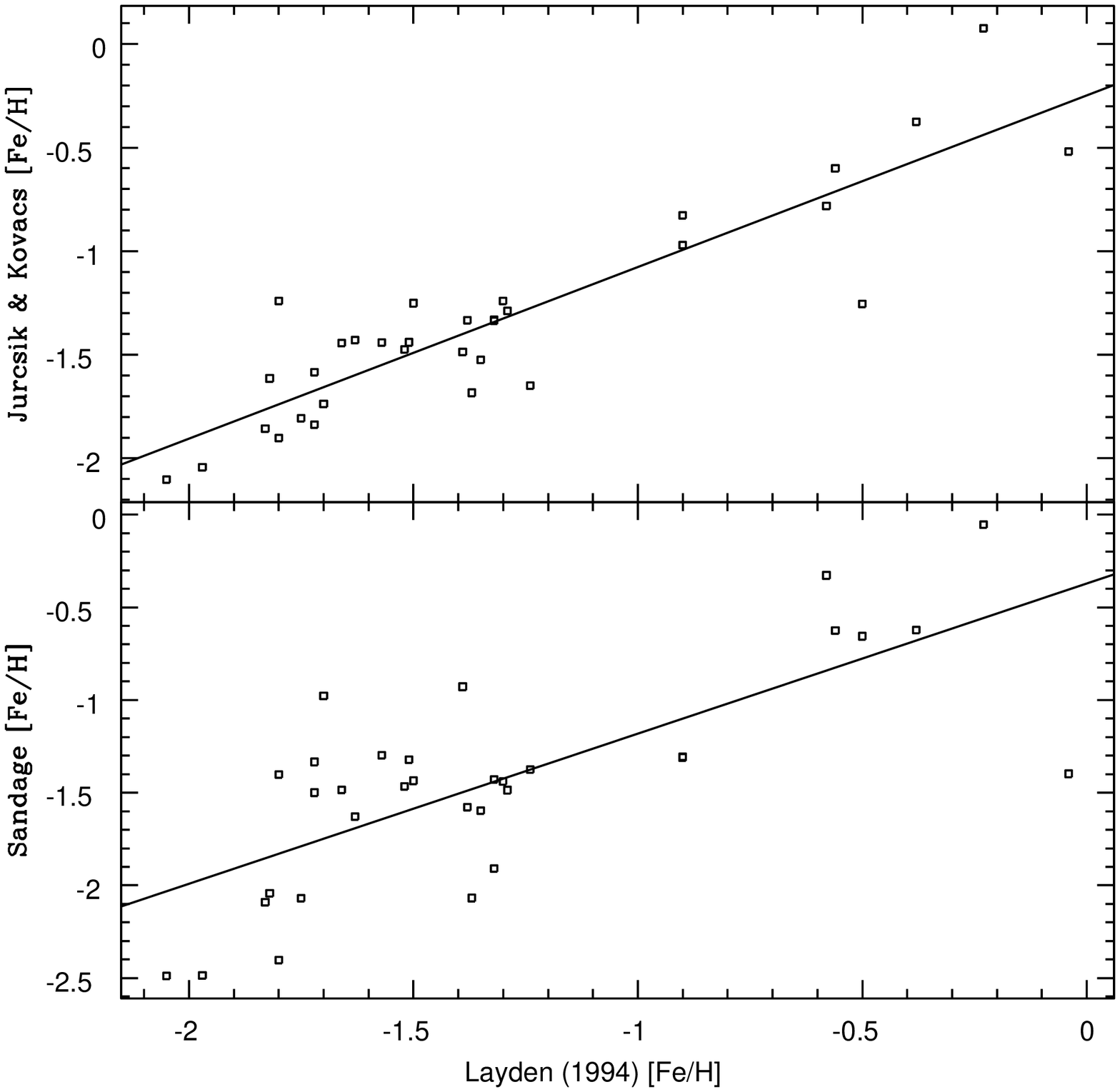}
\caption{A comparison between [Fe/H] values derived from both
photometric metallicity methods against the metallicity obtained
spectroscopically \citep{Layden:1994}.  In both plots the same 33 stars
were used in the comparison.  The upper plot is the comparison of the
[Fe/H] from the \citet{Jurcsik:96} method to the spectroscopic [Fe/H]
values.  The lower plot shows the comparison of the metallicity
derived from \citet{Sandage:2004}'s period-amplitude relation to the spectroscopic ones.}
\label{compare_feh}
\end{figure}

We compared the metallicity values for stars where [Fe/H] values were
derived from both methods.  In a few cases, large discrepancies
occurred; in a few instances the difference was as large as 3 dex.
These large differences were mainly due to a poor Fourier
decomposition of the light curve that did not have good phase
coverage.  For the larger discrepancies, we adopted the
metallicity derived from the Sandage method alone.  Nineteen stars had this
problem for our sample of RRab stars for which both metallicity methods were
used. Figure \ref{compare_feh} compares the two [Fe/H] solutions for a set of
33 RRab stars against \citet{Layden:1994}'s spectroscopically
determined metallicities.  These 33 stars were common to the work of
both \citet{Jurcsik:96} and \citet{Sandage:2004} and had good quality
light curves in our sample.  

If there was adequate agreement between the [Fe/H] values from both
methods, we calculated a weighted average of the metallicity.  We
adopt this weighted average as our ``best estimate'' photometric
metallicity for that star.  If a metallicity solution was not
determined from the Jurcsik \& Kovacs method, the result from
Sandage's method was our best estimate solution.  Table \ref{rrab_feh} lists
our best estimate [Fe/H] for our sample of RRab stars.

The distances derived for the NSVS RRab stars depended on an adopted
$M_{V}-[Fe/H]$ relation.  With the ``best estimate'' metallicity on
hand, we used the $M_{V}-[Fe/H]$ relation provided by
\citet{Cacciari:2003} to obtain the absolute magnitudes.  The
\citet{Cacciari:2003} relations are reproduced here in Equations
\ref{cacciaridist1} and \ref{cacciaridist2}.  The dust maps of
\citet{Schlegel:1998} were used to derive the extinction, assuming
$A_{V} = 3.1E(B-V)$.  For our sample of RRab stars, we assume most of
the stars are above the Galactic plane, and thus are not heavily
reddened.  For some of the analysis of the sample, we imposed a limit
of Galactic latitude ($|b| > 12^{\circ}$) to avoid any heavy reddening
that may be due to the Galactic plane. 

\begin{equation}\label{cacciaridist1}
M_{V}  =  (0.17 \pm 0.04)[Fe/H] + (0.80 \pm 0.10)  \qquad\mbox{for}\qquad  [Fe/H]  <  -1.5 
\end{equation}
\begin{equation} \label{cacciaridist2}
M_{V}  =  (0.27 \pm 0.06)[Fe/H] + (1.01 \pm 0.12)  \qquad\mbox{for}\qquad  [Fe/H]  >  -1.5
\end{equation}

Table \ref{rrab_properties} provides a listing of parameters of the NSVS RRab
candidates.  The NSVS identification number, equatorial coordinates, period,
amplitude scaled to the Johnson V system, and the intensity-weighted
mean magnitude of each star are listed.  We note that \citet{Wils:2006}
have created an independent catalog of RR Lyrae stars in the NSVS.  Of
the 1188 RRab candidate stars included in our Table
\ref{rrab_properties}, 568 are also included in the Wils et
al. catalog.  The differences in our catalogs mainly stem from the
selection criteria used and the duplication of object entries in the
NSVS database.  We discuss the differences in the two catalogs in
detail in section 4.

\subsubsection{Blazhko Effect and Photometric Metallicity}
The Blazhko Effect is a long term (typically tens of days) period
modulation of the primary light curve.  About 20 or 25\% of the RRab
variables in the solar neighborhood exhibit the Blazhko Effect to some
degree \citep{Szeidl:76, Smith:95}.  We thus need to
consider how unrecognized Blazhko Effect stars might influence our
derived metallicity.

Some, but probably not all, of the Blazhko variables are excluded by
restricting the application of the Jurcsik \& Kovacs method to stars
having formal uncertainties of less than 0.3 in the $\phi_{31}$
parameter.  For those Blazhko stars that slip past this constraint,
the Blazhko modulation of the light curve adds an extra uncertainty to
the derived value of [Fe/H].  This can result in the calculated [Fe/H]
to be either too high or too low, typically by 0.1 or 0.2 dex.  In 
\citet{Kovacs:2005}'s recent analysis of Fourier decomposition
metallicities for a sample of field RR Lyrae stars, he found a few Blazhko
variables among the outliers that had deviations in [Fe/H] of more than
0.4 dex from the spectroscopic values.  However, he suggests that,
although the issue needs more study, the average light curves of most
Blazhko variables may yield Fourier decomposition metallicities in
good agreement with spectroscopic determinations.

The Blazhko Effect can also influence metallicities calculated from
the period-amplitude diagram.  The Blazhko Effect acts to reduce the
amplitude of the RRab stars \citep{Szeidl:76, Szeidl:88, Smith:95}.
Typically, this can result in the observed amplitude of our light
curve being 0.1 magnitude less than for an equivalent non-Blazhko
star. The resultant value of [Fe/H] will thus be spuriously low by
about 0.15 when Sandage's calibration is employed.

\section{Discussion of ab-type RR Lyrae Results}

The source of the RRab sample in this work primarily comes from the
2MASS correlated version of the NSVS database.  Aside from the constraints on coverage mentioned earlier, the NSVS database
should provide a complete catalog of field RR Lyrae stars within the
solar neighborhood down to $V \sim 15$.  As a check, we took an early
RRab candidate list that contained 1188 stars and cross correlated it
with the GCVS \citep{Kholopov:1996}.  We took into consideration that
some of the stars in the GCVS have poor astrometric positions, with
uncertainties up to 1 arcminute.  To match the NSVS RRab candidates
with the GCVS RRab stars, our only option was to perform a positional
match.  A search grid with sizes varying from 2 arcminutes to 3
arcseconds was used. At the match limit of 3 arcseconds, only 133 NSVS
RRab stars were found in the GCVS. Thus, out of 1188 NSVS RRab
candidates, almost 90\% were not found in the GCVS catalog.  For a 2
arcminute positional match, 312 matches occurred, and an 1 arcminute
search grid yielded 291 matches. Thus, it is safe to conclude that at
least three quarters of the NSVS RRab candidates are not in the GCVS. 

We also did a reverse search of the northern GCVS RRab stars
in the NSVS database that had been correlated with the 2MASS catalog.
We expected that all bright GCVS RRab stars with declination greater
than $-30^{\circ}$ should be in the NSVS database.  We also selected
GCVS stars that were between 8th and 14th magnitudes in V.  In this reverse
search, 99 GCVS stars were not recovered in our working sample of RRab 
stars.  To investigate these missing RRab stars, we searched further in 
the 2MASS correlated NSVS database, as well as in the original NSVS
database located at Los Alamos.  All but seven of the GCVS stars could
be found in the LANL version of the database.  The reasons for the absence
of these seven variable stars (HP Aqr, AX Del, SZ Leo, DI Leo, NQ Lyr, 
AK Pup, and FI Sge) is unknown at this time.

The search of the 2MASS correlated database recovered 63 of the 99 missing
variable stars.  Forty-three GCVS stars were excluded because their entries
had fewer than 20 good observations.  Nine other missing stars had a 
Galactic latitude within $12^{\circ}$ of the plane.  As noted in \citet{Wozniak:2004a}, some confusion can exist in the identification of individual stars 
near the Galactic plane because of crowding.  Eleven stars were excluded
due to the selection criteria applied to periods, colors, and amplitudes in
identifying RRab candidates as described in section 3.1.  

\citet{Wils:2006} have recently completed a catalog of RR Lyrae stars
in the NSVS which was done independently of the present survey.  They
identified 785 probable RR Lyrae stars, including 188 that were
previously unknown.  Of the 785, 712 are of RRab type, about 468 of
which are brighter than magnitude 14. \citet{Wils:2006} also provide a
convenient table for cross identifying known RR Lyrae stars with those
in the NSVS.

In comparing our RRab candidates with those in \citet{Wils:2006}, we
find that most of the Wils et al. stars are also in our list, but not
all.  We also note that, because of the synonym problem in the NSVS,
the stars in our Table \ref{rrab_properties} are occasionally listed under different
NSVS identification numbers than the corresponding variables in
\citet{Wils:2006}.  There are some 60 RRab stars in the Wils et
al. list that are brighter than 14 magnitude that are not included in
Table \ref{rrab_properties}.  At least 50 of these appear to be genuine RRab stars
that were excluded by one of the selection criteria we used.  Most
often the period criteria caused the omission of the star from our
sample.  For the remaining 10 stars there is still some uncertainty as
to whether an RRab classification is correct.  For consistency, we
have not included the additional Wils et al. stars in our analysis,
but have retained the original sample from our search criteria.

\subsection{Period Distribution}

\begin{figure}
\figurenum{10}
\includegraphics[angle=-90,scale=0.6]{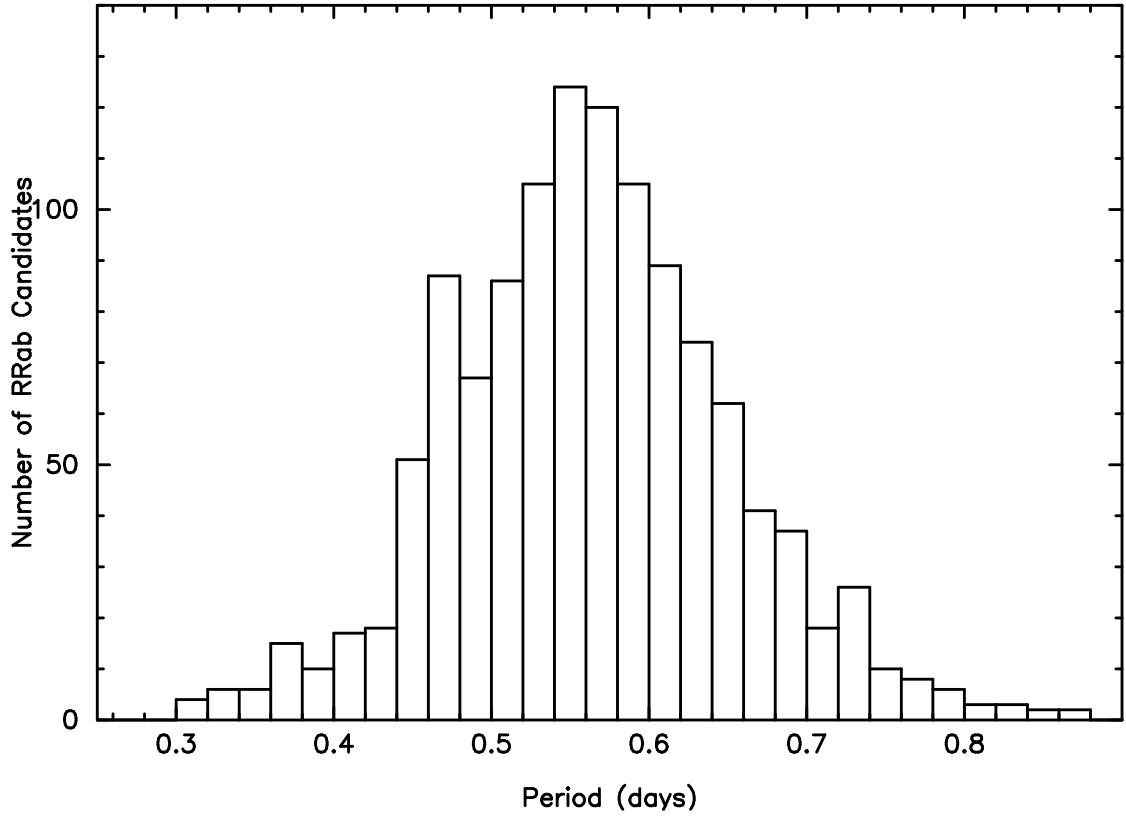}
\caption{The period distribution of 1188 NSVS RRab stars.  The average
  period of these stars is $0.563 \pm 0.001$ days.}
\label{rrabhist}
\end{figure}

The period distribution of 1188 NSVS RRab stars was examined through a
period histogram, shown in Figure \ref{rrabhist}.  In this histogram, the
average period of the RRab stars is $0.563 \pm 0.001$ days.  This mean
value is close to the average period of 0.55 days seen for RRab stars
of Oosterhoff type I globular clusters, and is much shorter than the
0.64 days typically seen in clusters of Oosterhoff type II
\citep{Smith:95}.  However, it is clear that the distribution of
periods among the field RRab stars is wider than that seen within
individual Oosterhoff type I clusters.  Instead, the field RRab period
histogram is a consequence of the mixture of several RR Lyrae
populations, as can be seen from the distribution of the RRab sample
with respect to the Galactic plane.

The outer halo star studies of \citet{Carney:1996} and cluster studies
of \citet{Lee:1999} hypothesized different origins for Oosterhoff type
I and Oosterhoff type II clusters.  \citet{Lee:1999} used a sample of
125 RRab stars within 3 kpc of the plane and 61 RRab stars farther
than 5 kpc from the plane to examine the relative distributions of
halo RRab stars belonging to Oosterhoff group I and Oosterhoff group II.
\citet{Lee:1999} concluded that there was evidence that field RRab
stars associated with an Oosterhoff II population were more likely to
be found in the low Z group, being rarer at $|Z| > 5$ kpc.

\begin{figure}
\figurenum{11}
\includegraphics[angle=-90,scale=0.6]{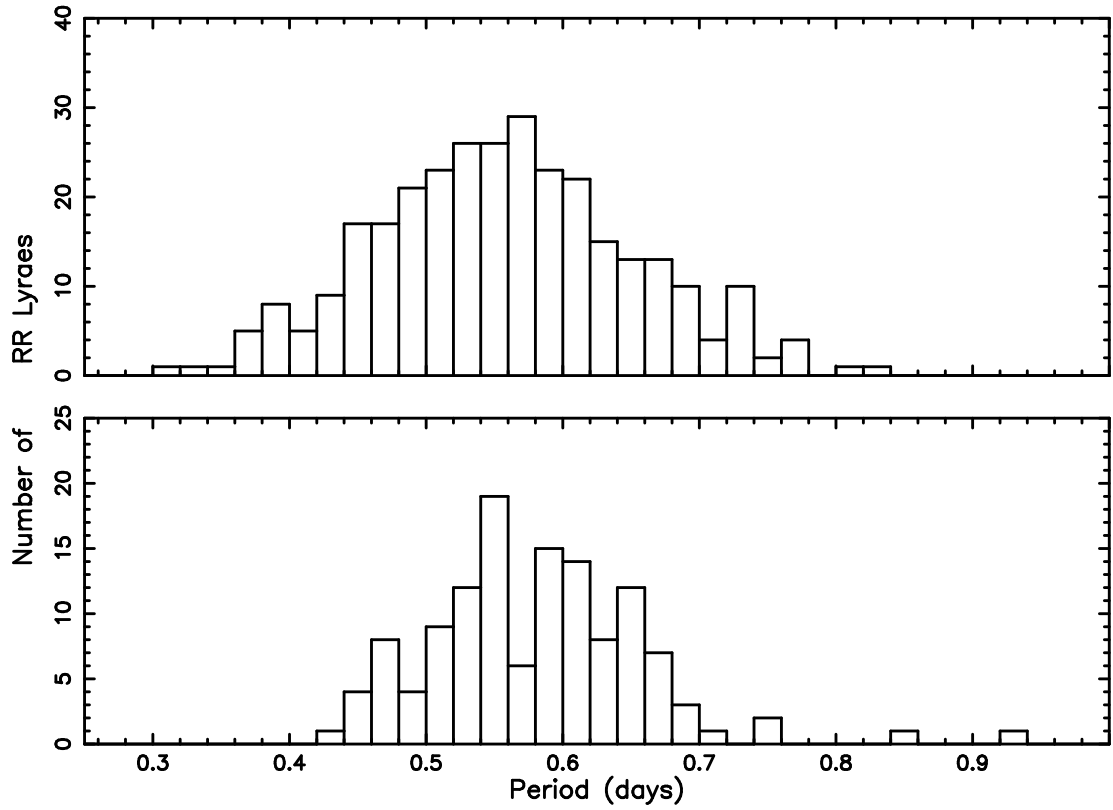}
\caption{The period distribution, as discussed in section 4.1,
  separated into two groups defined by the star's $|Z|$ distance from
  the Galactic plane.  The upper plot is for RRab stars found at $|Z|
  < 2.0$ kpc.  The average period for those stars is 0.557 days.  The
  lower plot shows the period distribution for RRab stars found at
  $2.0 < |Z| < 5.0$ kpc.  The average period for this distribution is
  0.580 days.}
\label{rrabhist_z}
\end{figure}

Because our sample of RRab stars becomes seriously incomplete as we
move to a distance beyond 5 kpc, we cannot extend our survey to 
distances as large as those studied by \citet{Lee:1999}.  However, we
can investigate trends in RRab populations within 5 kpc of the plane.
The RRab sample was divided into regions defined by the distance
from the Galactic plane, Z.   Figure \ref{rrabhist_z} shows the period
distribution for RRab stars found close to the plane,
$|Z| < 2$, kpc and at larger distances, $2 < |Z| < 5$ kpc. The average
period of RRab stars found close to the plane shows a mean period
indicative of an Oosterhoff I class.  The average RRab period for the
region $|Z| < 2$ kpc was $0.557 \pm 0.001$ days.  For the regions
slightly further away, the average period shifted to an intermediate
value between the Oosterhoff I and II types.  The average RRab period
for this group of stars in these regions was $0.580 \pm 0.001$ days.
It is clear, however, that the mean period in the low $|Z|$ sample is
strongly influenced by the presence of a short period disk population
of RRab stars.  To winnow these from the sample requires a more
detailed consideration of the RRab properties, which we discuss in the
next section. 

\subsection{Period-Amplitude Trends}

A Bailey, or period-amplitude, diagram can be used as a
diagnostic tool to investigate the Oosterhoff classification of RRab
stars.  \citet{Preston:59} discovered that the location of field RRab
stars in the Bailey diagram was a function of metallicity.
\citet{SKS:1981} noted that the same effect among the RRab stars of
globular clusters might be explained if metal-poor RRab stars were
more luminous than their metal-rich counterparts.
\citet{Clement:2000} furthered the argument that RRab stars of
Oosterhoff type I and II occupy separate and distinctive lines in the
Bailey diagram.

We will use the linear trends from \citet{Clement:2000}, which are
based on the RRab stars found in M3 (Oosterhoff I type globular
cluster) and $\omega$ Cen (an Oosterhoff II cluster) as representative
of the period-amplitude relations of Oosterhoff I and II clusters.  M3,
at [Fe/H]$= -1.5$, is one of the most metal-poor of the Oosterhoff I
clusters.  Thus, we would expect field RRab stars with Oosterhoff I
properties to lie along or to the left of the M3 locus in the Bailey
diagram.  We would expect RRab stars of Oosterhoff type II to lie
along, or perhaps slightly to the right, of the $\omega$ Cen locus in
the Bailey diagram.

\begin{figure}
\figurenum{12}
\includegraphics[angle=-90,scale=0.6]{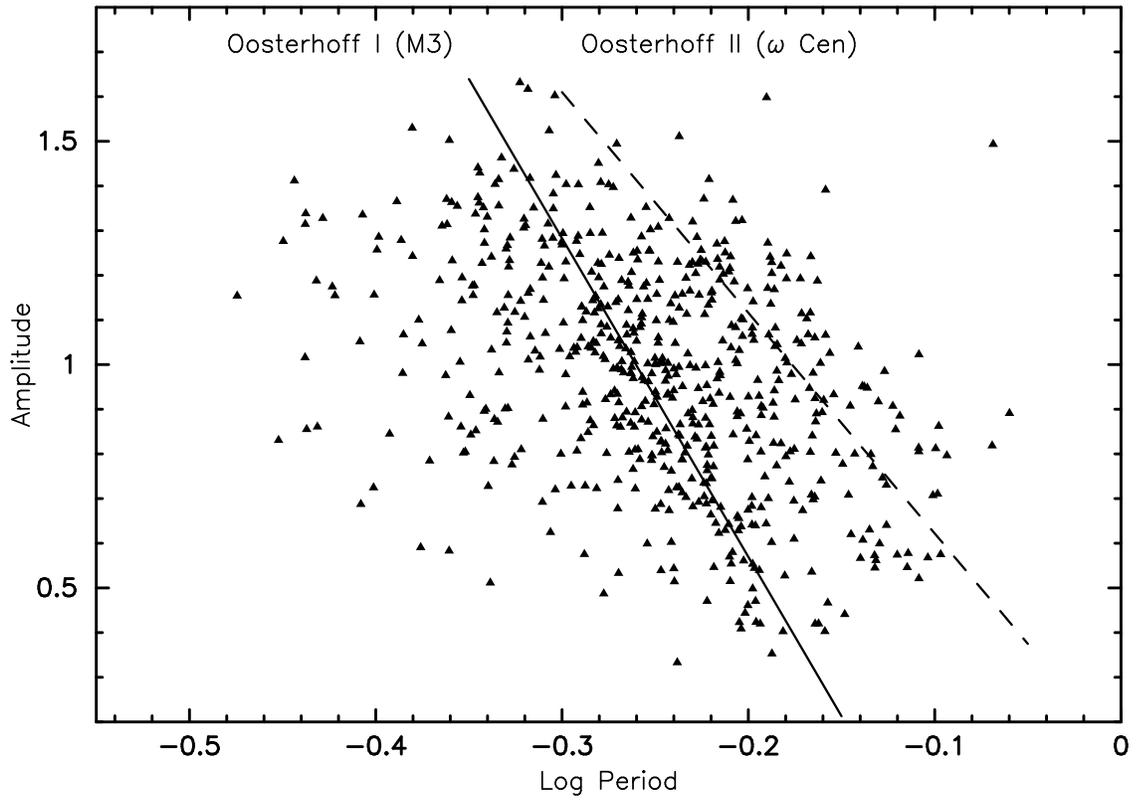}
\caption{The NSVS RRab stars plotted in a Bailey, or period-amplitude,
  diagram, as discussed in section 4.2.  \citet{Clement:2000}'s
  Oosterhoff relations are overplotted here.}
\label{padiag_rrab}
\end{figure}

Figure \ref{padiag_rrab} is our period-amplitude diagram for the 608
stars for which we had determined the period and amplitude.
In this Figure we also overplot the linear Oosterhoff relations of
\citet{Clement:2000}.  The RRab stars do not clearly
fall along one or the other relation, as is usually seen in the cases
for globular clusters.  However, it appears that a majority of the
stars fall near the Oosterhoff I relation, which is in agreement with
the result from the period histograms.  \citet{Cacciari:2005} obtained
a mean period-amplitude relation for RRab stars in M3 that is somewhat
different from that in \citet{Clement:2000}.  In particular, Cacciari
et al. noted that the more numerous regular and the rarer, more evolved
M3 RRab stars occupied somewhat different positions in the
period-amplitude diagram.  The Cacciari et al. trend line for the
regular (less evolved) RRab stars is generally similar to that of
Clement and Rowe, but is slightly displaced and includes a quadratic
term, so that the line flattens slightly at large amplitudes.  Use of
the Cacciari et al. rather than the Clement and Rowe Oosterhoff I line
would not, however, change any of the conclusions we arrive at below.
Even with the revised trend lines, significant populations exist near
the Oosterhoff II trend line and at short periods to the left of the
M3 Oosterhoff I line. 

\subsection{Oosterhoff Dichotomy Classification for NSVS RRab Stars}

\begin{figure}
\figurenum{13}
\includegraphics[angle=-90,scale=0.6]{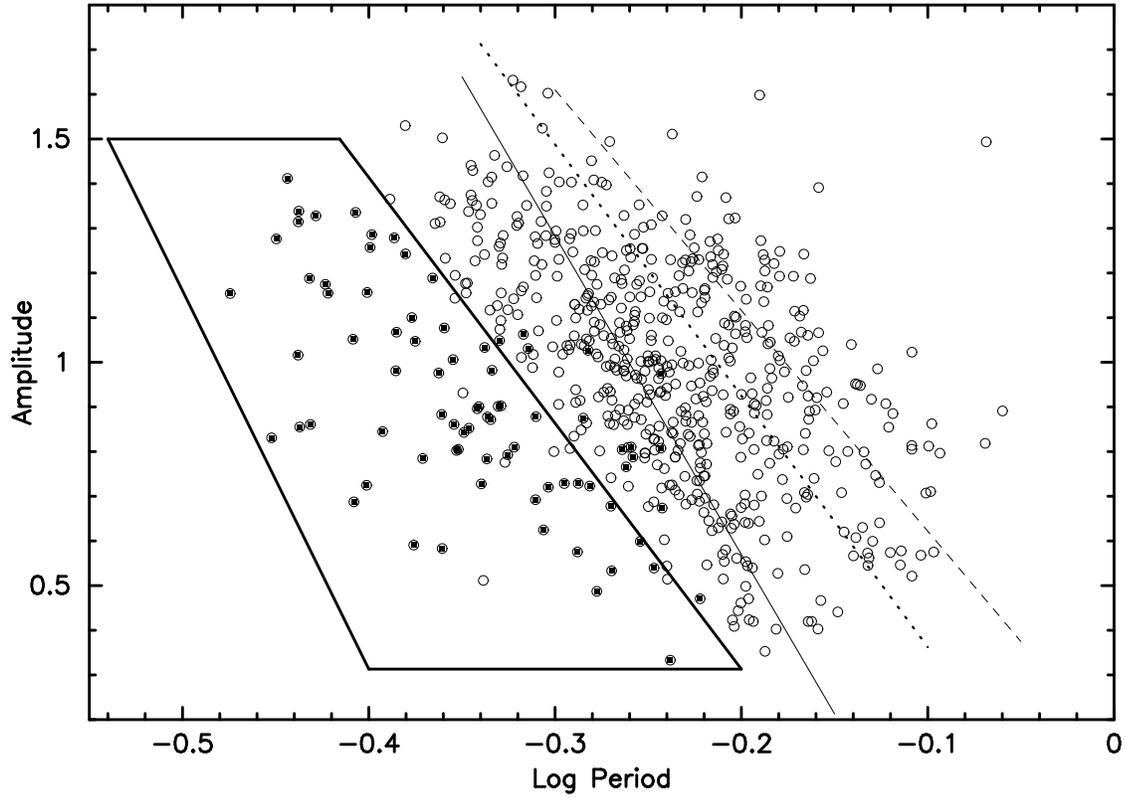}
\caption{The Bailey diagram for our RRab sample is revisited here, but
  now taking a closer look at the metal-rich sample, as discussed in
  section 4.3.  The RRab stars with $[Fe/H] > -1$ are identified by a
  filled circle. Again, \citet{Clement:2000}'s Oosterhoff trends are
  overplotted as the solid and dashed lines.  The dotted line is our
  arbitrary division between the Oosterhoff I and II groups.  The box
  is an aide to show where the majority of the metal rich RRab stars
  are found in this diagram.}
\label{pa_oo}
\end{figure}

A more careful classification for the NSVS RRab stars can be performed
with the addition of metallicity information to the period
and amplitude.  Three groups of stars are identified in our
period-amplitude diagram (Figure \ref{pa_oo}): the Oosterhoff I RRab
stars, the Oosterhoff II RRab stars, and short period group of RRab stars.
Figure \ref{pa_oo} is a reproduction of Figure \ref{padiag_rrab} but with the
Oosterhoff groups identified.  The short period stars in the ``metal rich'' box of
Figure \ref{pa_oo} were scrutinized for RRc stars contaminating the RRab
sample.  Almost all of these short period stars were found to be 
genuine ab-type RR
Lyrae stars with metallicities of [Fe/H] $> -1$.  The right edge of this
box is located at $[Fe/H] = -1$, as defined by the Sandage metallicity relation
(Equation \ref{ampmet}).  All the stars with the filled circles were
identified with a metallicity richer than [Fe/H] $= -1$, according to
the best estimate metal abundance in Table \ref{rrab_feh}.

\begin{figure}
\figurenum{14}
\includegraphics[angle=-90,scale=0.6]{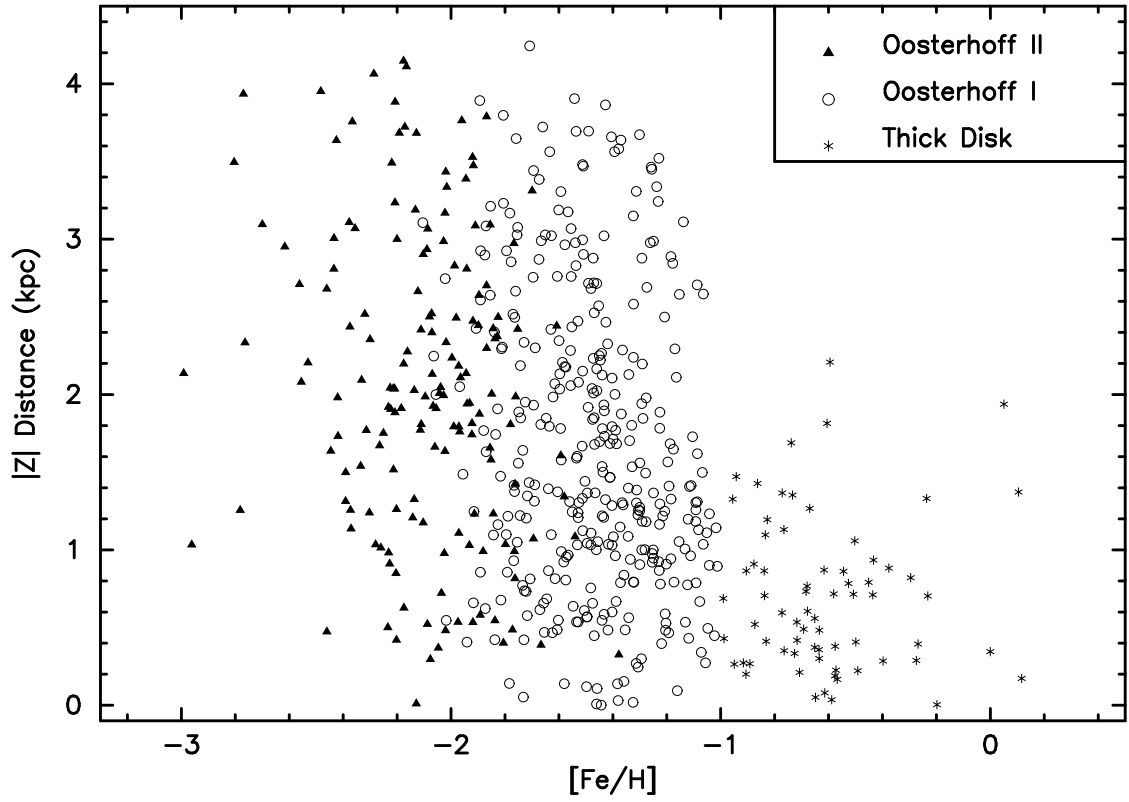}
\caption{Distribution of the RRab sample, grouped according to the
  Oosterhoff group as discussed in section 4.3, plotting metallicity against
  $|Z|$.  The metallicity is the best estimate value listed in Table
  \ref{rrab_feh}.}
\label{feh_z}
\end{figure}

\begin{figure}
\figurenum{15}
\includegraphics[angle=-90,scale=0.6]{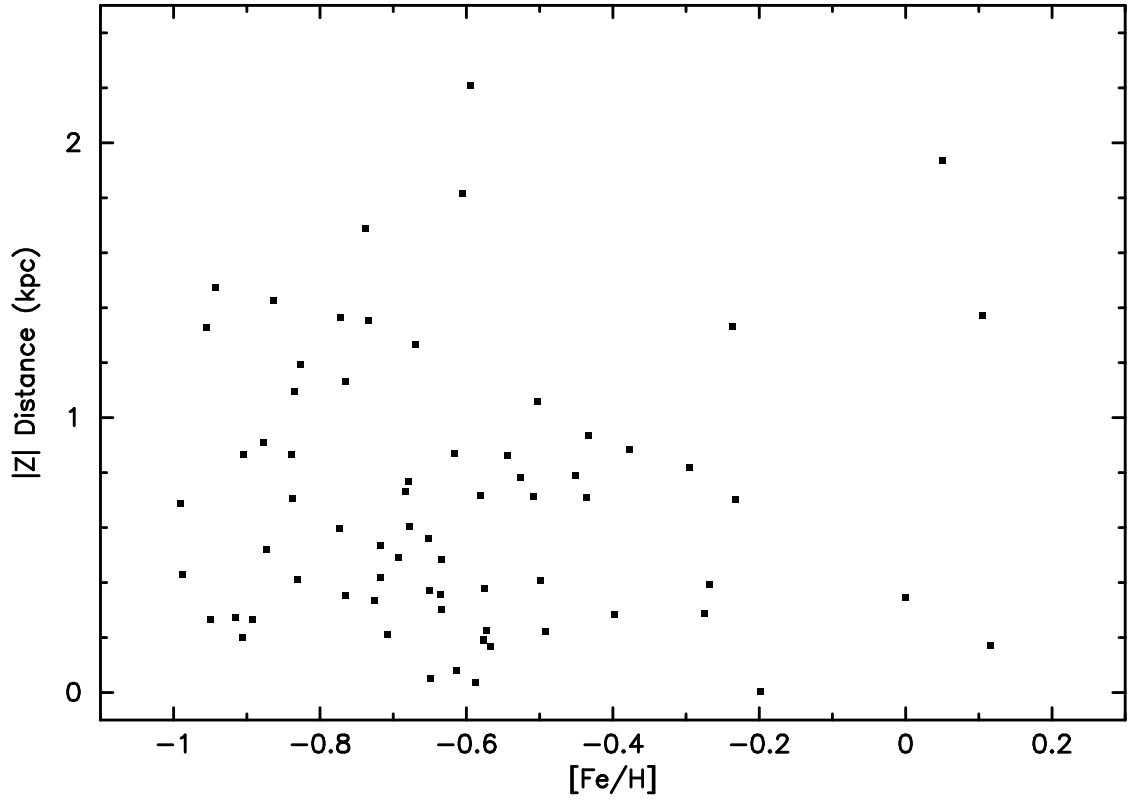}
\caption{A close up of the metallicity distribution of the metal rich
  group in $|Z|$ distance.  The metallicities are again the best
  estimate value.  The six stars appearing above the bulk of stars had
  metallicities only derived from \citet{Sandage:2004}'s method.}
\label{tdmetz}
\end{figure}

To examine the spatial distributions of the two Oosterhoff groups and
the metal-rich group of RRab stars, we plot in Figure \ref{feh_z} the
[Fe/H] distribution versus the distance from the Galactic plane, $|Z|$.  In
Figure \ref{tdmetz}, we see that the metal rich stars with [Fe/H] $>
-1$ all lie close to the Galactic plane.  The metal poor stars exhibit an 
extended distribution more consistent with the existence of a component
with halo properties.  However, below we will argue that a fraction of
the RRab stars more metal-poor than [Fe/H] $= -1$ belongs to a thick
disk population.  Due to the faintness limit we imposed for our sample
of RRab stars, our RRab stars probe to about 4.5 kpc away from the
Galactic plane. 

\subsubsection{Oosterhoff I and II}

We divide the RRab stars in Figure \ref{pa_oo}, excluding the metal
rich stars confined in the box, along the dotted line.  Stars to the
left of this line are denoted as belonging to Oosterhoff I, while
those to the right are credited to Oosterhoff II.  Ideally, we might
want to draw the dotted line to correspond to a metal abundance of
[Fe/H]$ =-1.7$, which is the approximate boundary between globular
clusters of Oosterhoff types I and II.  However, if we use Sandage's
amplitude-log period-[Fe/H] relation to identify the location of the
[Fe/H]$ = -1.7$ boundary, the dividing line intersects with the
Oosterhoff I relation of \citet{Clement:2000}.  This may suggest a
small problem either with Sandage's amplitude-period-metallicity
calibration, or with the adopted M3 trend line.  

\subsubsection{Metal Weak Thick Disk Population?}

\begin{figure}
\figurenum{16}
\includegraphics[angle=-90,scale=0.6]{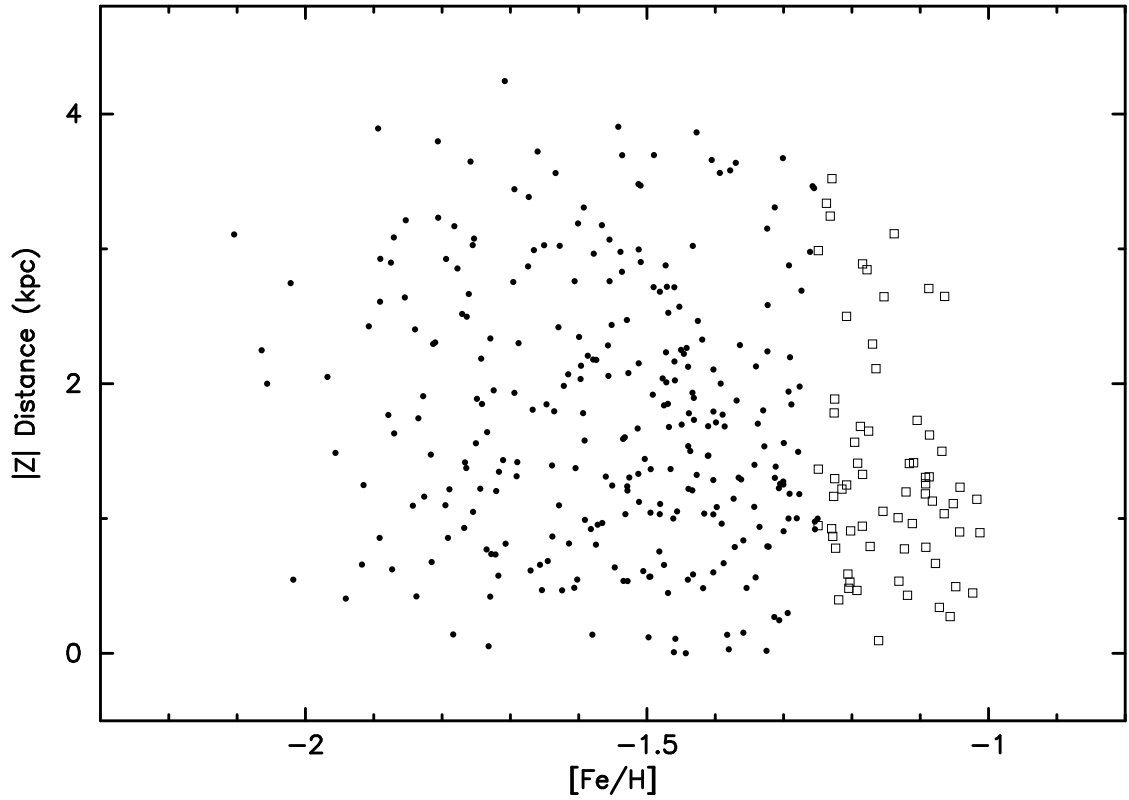}
\caption{Distribution of the Oosterhoff I RRab stars.  The open squares
are those Oosterhoff I RRab stars with $[Fe/H] > -1.25$.  The filled
circles are the metal poorer Oosterhoff I RRab stars with $[Fe/H] <
-1.25$.  Note the slight clump of RRab stars near $|Z| = 1$ kpc for
the metal richer RRab stars.}
\label{oo1subgroup}
\end{figure}

The Oosterhoff I candidates were separated into two subgroups with the
division occurring at $[Fe/H] = -1.25$.  Note in Figure \ref{oo1subgroup}
the existence of a relatively large number of stars of [Fe/H] $>
-1.25$ within 2 kpc of the Galactic plane.  
To investigate the implications of this group of stars, we first
consider whether the apparent overabundance of disk stars in the more
metal-rich Oosterhoff I group might reflect a bias in the
identification of RRab stars.  We used the more metal-poor Oosterhoff
II sample as a control, assuming that it represents a pure or nearly
pure halo population.  The detection probabilities of the Oosterhoff
II RRab stars are similar to those of Oosterhoff I RRab of comparable
amplitude within our survey.  We compared the number of Oosterhoff I
and II type stars at different $|Z|$ distance intervals.  Table
\ref{numratio} summarizes our calculated number ratios of these stars.

We assume that the two $|Z|$ distance intervals listed in Table
\ref{numratio} best span the disk and inner halo population of our
sample of stars.  The regions above 5 kpc are excluded because of our
adopted magnitude limit.  In the region closer to the plane ($|Z| <
2.0$ kpc), we also have an additional constraint that the RRab stars
have a Galactic latitude greater than $12^{\circ}$, i.e. $|b| >
12^{\circ}$.  This constraint was imposed to avoid the regions of
heavy interstellar extinction near the plane where the
\citet{Schlegel:1998} reddening maps are uncertain and where our
survey incompleteness becomes more serious.

Our number ratio calculations suggest that different Oosterhoff groups
are dominant in the two $|Z|$ regions.  In the case of Oosterhoff II
(Oo II) to Oosterhoff I (Oo I), the ratio shows that the Oo I stars
outnumber the Oo II stars 2 to 1 for the $2.5 < |Z| < 3.5$ kpc
region.  However, closer to the plane, the same ratio seems to
indicate that Oo I RRab stars increase more rapidly as we approach the
plane than is the number of Oo II RRab stars. This result is certainly
consistent with the presence of a disk component within our Oosterhoff
I group.  We repeat this analysis to investigate the disk and halo
populations, but divide the Oosterhoff I sample into two subgroups by
metallicity.  The metal richer Oosterhoff I subgroup ([Fe/H]$ >
-1.25$) and the metal poorer Oosterhoff I subgroup ([Fe/H]$ < -1.25$)
were each compared with the Oosterhoff II group at the same $|Z|$
distance intervals as in the above analysis.  Table \ref{oo_numbers}
provides the numbers of RRab stars assigned to each Oosterhoff group
at each $|Z|$ interval.  Table \ref{ooI_num} lists the number of RRab
stars that make up the two Oosterhoff I subgroups.

These number ratios suggest that the disk component of the RRab
sample likely extends to metal abundances somewhat lower than our
original [Fe/H] $= -1.0$ cutoff.  The numbers of stars in the samples
are not so great as to allow this conclusion to be drawn at more than
the two sigma level.  Full kinematic data on the sample stars would
allow this conclusion to be tested more fully.  Taking the ratios at
face value, however, we estimate that approximately 60\%
of our metal-rich Oosterhoff I stars in the region $|Z| <$ 2.0 kpc
belong to the thick disk.  There is even some evidence for a metal-weak
thick disk component \citep{Norris:1985, Morrison:1990, Twarog:1994,
  Beers:1995} but the sample is not large enough to make a completely
convincing case.

We note that uncertainties in our derived values of [Fe/H] can scatter
metal-rich stars into our metal weak sample and vice versa.  The fact
that there are very few stars in our [Fe/H] $> -1$ sample that also
have $|Z| > 2$ kpc indicates that this effect is not large and does
not account for the overabundance of stars near the disk in our
Oosterhoff I metal-rich group.  Such an effect might, however,
contribute to the smaller excess near the disk among the Oosterhoff I
metal-poor group.  In order to clearly identify whether a star belongs
in the thick disk or halo population, the kinematics of the star must
be known.  The kinematic information can provide stronger evidence for
testing whether there exists a thick disk population in our Oosterhoff
I group of stars.  A program to obtain full space motions for these,
few of which now have known radial velocities, would certainly be of value.

We note that there is a continuum of properties among these field RRab
stars. The stars with higher metal abundances as determined by the Fourier
parameter method or (when such are available) by Layden's (1994)
spectroscopy, lie to the left in the period-amplitude diagram, whereas
the more metal-poor stars lie to the right.  The apparently metal-rich
RRab stars in the unusual globular clusters NGC 6388 and NGC 6441
\citep{Layden:1999, Pritzl:2001,Pritzl:2002, Pritzl:2003, Clementini:2005}
break this trend: falling to the right in a period-amplitude diagram
although NGC 6388 and NGC 6441 appear to be about as metal-rich as 47
Tucanae.  Thus, there is some similarity in the structure of the RRab
stars in the solar neighborhood field, both metal-rich and metal-poor,
that may extend to most but not all of their globular cluster
counterparts.

\subsection{Thick Disk Stars}

\begin{figure}
\figurenum{17}
\includegraphics[angle=-90,scale=0.6]{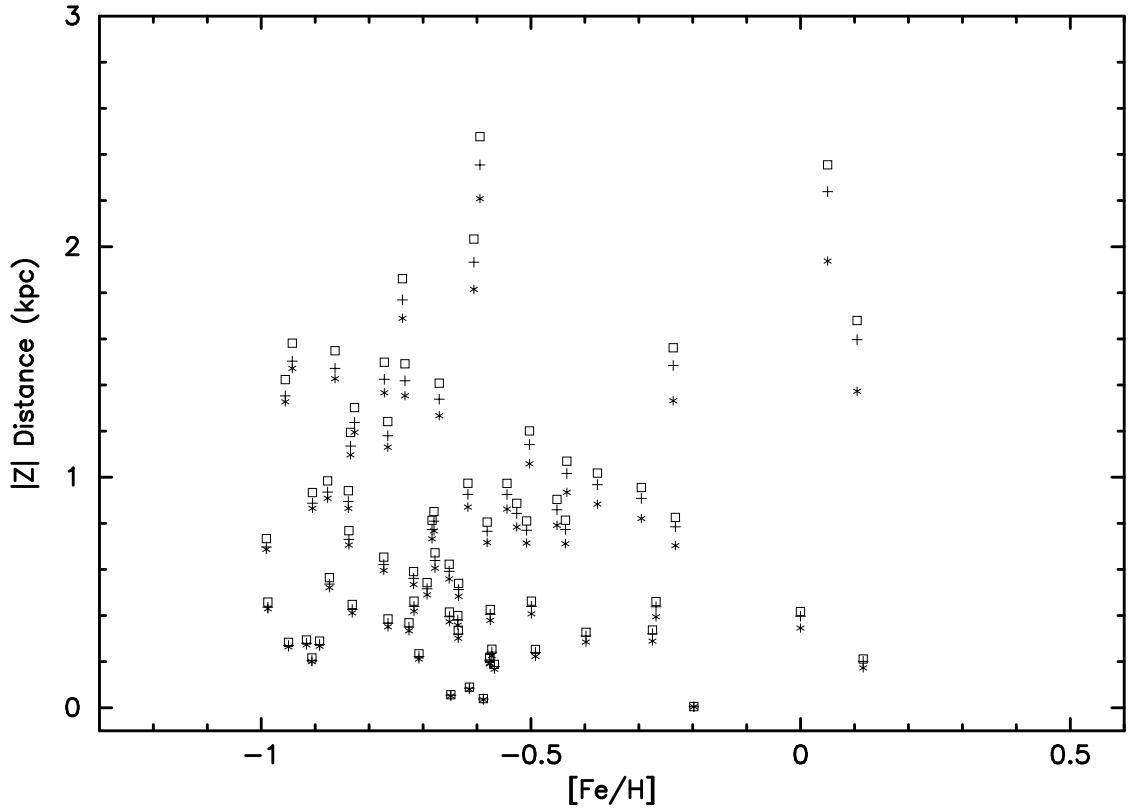}
\caption{Metal-rich thick disk sample trend with $|Z|$ distance.  As
  described in section 4.4, we checked the trend observed in our metal
  rich group to see if it was an artifact of our choice of
  $M_{V}-[Fe/H]$ relation \citep{Cacciari:2003}.  The asterisks are
  for distances determined from this relation.  The open squares are
  for distances derived with $M_{V} = 0.6$ \citep{Smith:95}, and the
  crosses are for distances derived with $M_{V} = 0.71$
  \citep{Layden:1996}.}  
\label{td_distfeh}
\end{figure}

The metal rich star distribution in Figure \ref{tdmetz} shows that most of
these stars are close to the Galactic plane.  The distribution of
these stars with respect to $|Z|$ is broadly consistent with that
expected of a thick disk population.  However, the distribution of these stars
appears to be a function of metal abundance.  Although there are a
few exceptions, there is a trend toward lower $|Z|$ distance as [Fe/H]
goes from -1 to 0.  This may indicate the existence of a thin, or at
least a less thick, disk component among the most metal-rich of
the RRab stars.  To check whether this thick disk trend might be an
artifact of our choice of $M_{V}-[Fe/H]$ relation, we rederived the
distances using fixed $M_{V}$ values (+0.6 and +0.71, respectively
\citet{Smith:95, Layden:1996}).  Regardless of how we arrived at the
distances, the trend was still present, as can be seen in Figure
\ref{td_distfeh}.  Thus, the tendency toward decreasing $|Z|$ with
increasing [Fe/H] does not seem to be imposed on the data by our
particular choice of $M_{V}-[Fe/H]$ relation.

\subsubsection{Scale Height of the Thick Disk}
Using the 589 RRab stars for which we are able to derive reliable
photometric metallicities, we have investigated the scale heights of
the thick disk and inner halo for different subgroupings within the
data.  To derive scale height, h(Z), of these groups we binned data
for each subgroup into an average of either 6 or 10 stars per $|Z|$
bin.  For the thick disk scale height calculation, we considered the
full sample as well as a subsample that excluded stars found close to the
plane ($|b| < 12^{\circ}$).  In the latter case, we also implemented a limit to
the volume in which we calculated the densities to be used in the
determination of the scale heights.  The adopted limits were $|b| >
12^{\circ}$ and a radius of 2.0 kpc from the Sun.  Therefore, the
shapes of the volumes were essentially frustums of a cone until we
reached the search radius limit of 2.0 kpc, which then reverted to the
volume of a cylinder.  A correction had to be applied to these volumes
for areas of the sky that were too far south to be included within the
ROTSE-I survey.  In Figure \ref{aitoff}, this region can be seen as a
hole where no RRab stars were detected for our sample.  This region is
bounded by Galactic longitude $240^{\circ} < l < 20^{\circ}$ and
Galactic latitude $-90^{\circ} < b < +30^{\circ}$.

\begin{figure}
\figurenum{18}
\includegraphics[scale=0.8]{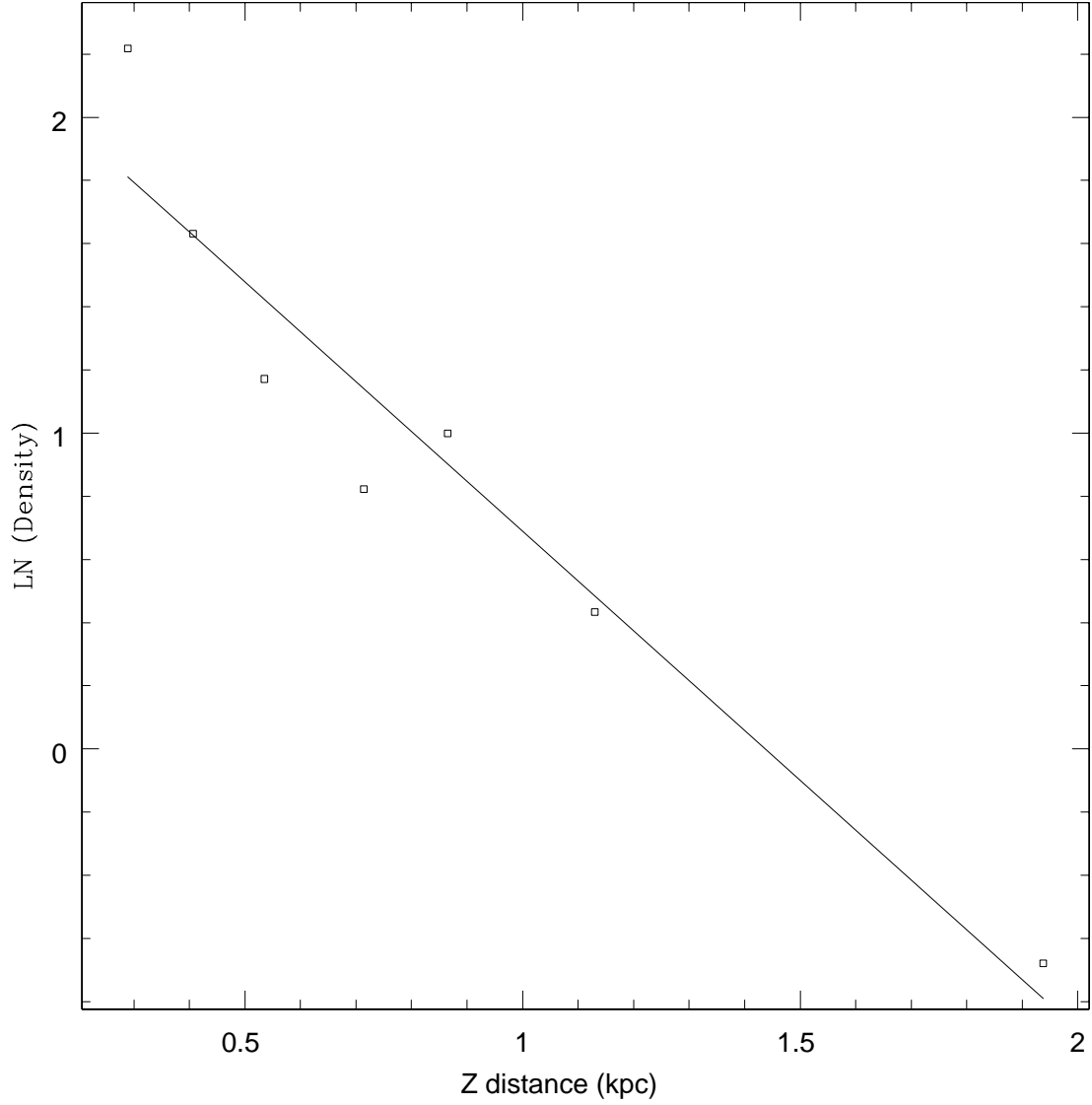}
\caption{The natural log of the density against the $|Z|$ distance for
  the case of binning 6 stars.  The corresponding plot for the scale
  height is Figure \ref{sclhgt6}.} 
\label{logdensity6}
\end{figure}

\begin{figure}
\figurenum{19}
\includegraphics[scale=0.8]{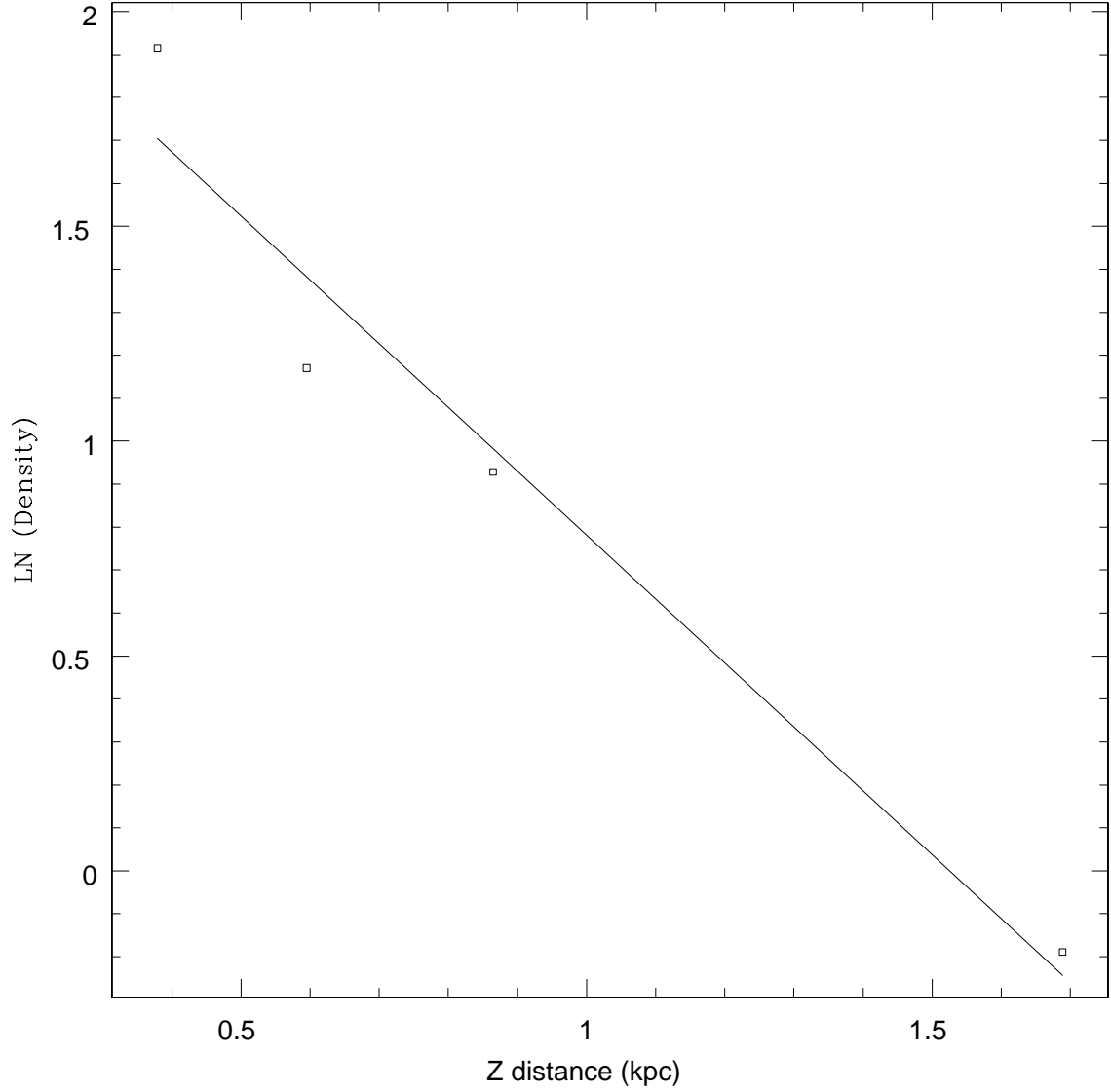}
\caption{This plot is similar to Figure \ref{logdensity6} but for the
  case of binning 10 stars per $|Z|$ bin.  The corresponding plot for
  the scale height is Figure \ref{sclhgt10}.}
\label{logdensity10}
\end{figure}

\begin{figure}
\figurenum{20}
\includegraphics[angle=-90,scale=0.6]{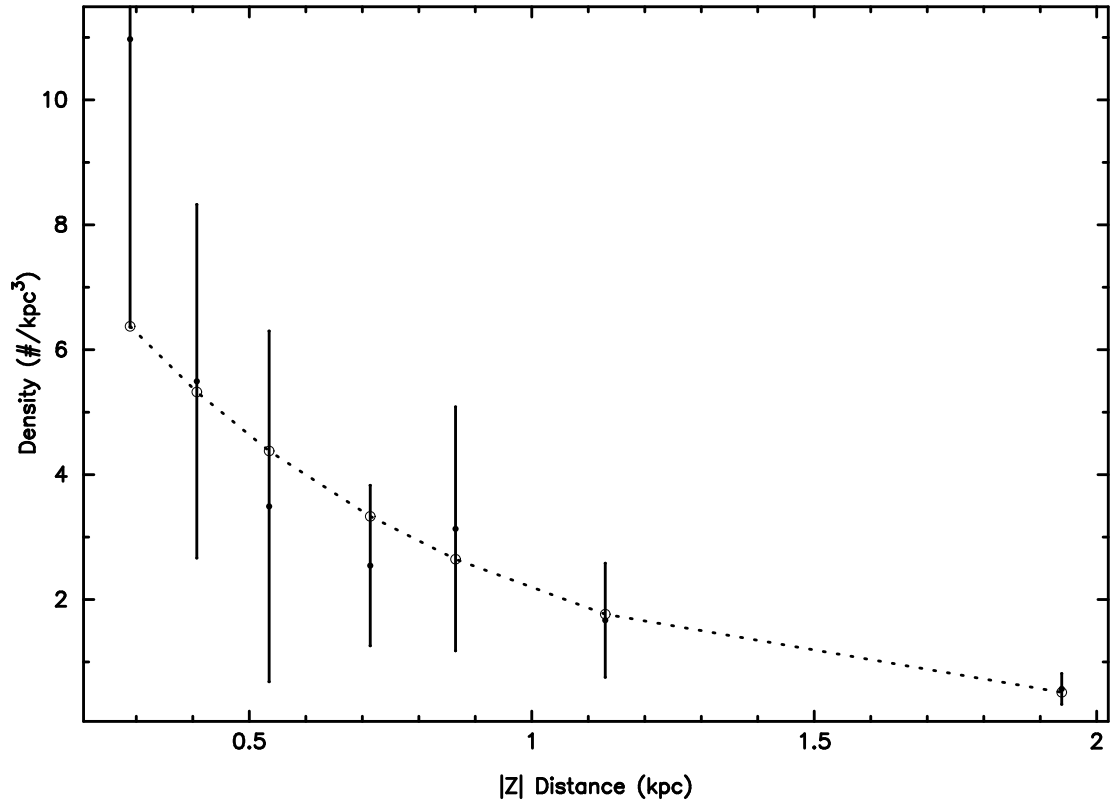}
\caption{RRab density distribution with respect to distance away from
  the Galactic plane, $|Z|$.  To calculate the scale height of the
  thick disk, a plot of the density of the stars per $|Z|$ bin was
  created.  In this plot, six stars are in each $|Z|$ bin.  The scale
  height for the metal rich thick disk is $0.66 \pm 0.16$ kpc.  We
  have accounted for the detection efficiency, as described in
  \citet{Amrose:2001}.} 
\label{sclhgt6}
\end{figure}

\begin{figure}
\figurenum{21}
\includegraphics[angle=-90,scale=0.6]{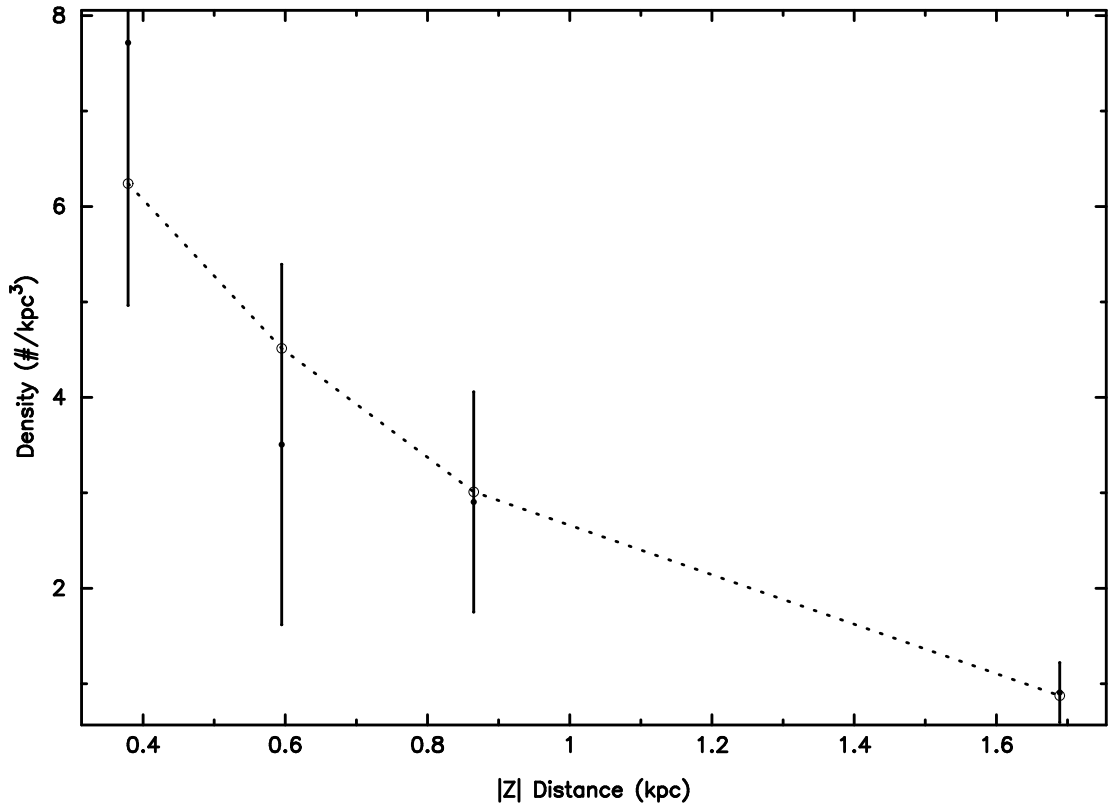}
\caption{This is the same as Figure \ref{sclhgt6} but we have binned 10
stars per $|Z|$ bin.  With detection efficiency accounted for, the
scale height is $0.67 \pm 0.17$ kpc.}
\label{sclhgt10}
\end{figure}

Once the densities were calculated, they were binned in $|Z|$, and an
exponential function was fitted to the results.  Figures
\ref{logdensity6} and \ref{logdensity10} show how the log density
falls off with the $|Z|$ distance bin for the full sample of
70 metal-rich stars. When the constraints on volume and $|b|$ are applied,
55 of these metal-rich RRab stars remain in the sample.  
For comparison, we also performed similar scale
height calculations for 330 RRab stars within the Oosterhoff I group
and for a ``halo'' sample of 428 RRab stars with $[Fe/H] < -1.25$.
The stars selected for these groups were subject to the same volume
limitations as those in the metal-rich group.  Although it would have
been interesting to separately determine scale heights for the
Oosterhoff II stars and the more metal-rich and metal-poor Oosterhoff
I groups, the numbers of stars within these subgroupings are
insufficient to give robust results when the volume constraints are
applied.  The halo sample results are, in any case, given merely
for completeness. The RRab sample we used does not
go deeply enough into the halo to yield a reliable value of scale
height.  Results are shown in Table \ref{sclhgt}.  Figures
\ref{sclhgt6} and \ref{sclhgt10} show plots of the density of the
metal-rich thick disk stars with $|Z|$, where the adopted error bars
are indicated.

The weighting of the density points is important to our scale height
solution.  If we apply no weighting, the calculated scale heights for
the metal-rich sample are much smaller, about 0.3 to 0.4 kpc, as
reported in \citet{Kinemuchi:2005}.  We believe that the scale height
calculation including weights depending on the uncertainties in
distance and Poisson statistics are the more reliable.

The scale height derived from the 6 stars per $|Z|$ bin case was $0.65
\pm 0.17$ kpc (Figure \ref{sclhgt6}).  For the 10 stars per $|Z|$ bin,
the scale height was $0.68 \pm 0.18$ kpc for our thick disk sample of
70 stars (Figure \ref{sclhgt10}).  These scale heights are smaller
than the canonical scale height of a pure thick disk sample ($\sim 1$
kpc).  Applying the volume constraints gives slightly greater, but also
more uncertain results, around 0.8 kpc.  In evaluating these results
we must consider whether there exists any selection bias against the
identification of RRab variables at large $|Z|$.  However, such a bias seems
unlikely since there is no large falloff in the detection efficiency
for RRab stars in the NSVS sample until one reaches $V > 14$, which we
set as the limiting magnitude for our RRab sample.  Application of the
efficiency corrections as a function of apparent magnitude found by
\citet{Amrose:2001} produces only a small change in the calculated
metal-rich group scale height, yielding $0.66 \pm 0.16$ kpc for the 6
stars per bin solution and $0.67 \pm 0.17$ kpc for the 10 star per bin
solution.  The correction from \citet{Amrose:2001} is an upper limit
on the actual correction expected for the stars used in calculating
our scale heights since the  number of observed data points in this
study is often greater than in the \citet{Amrose:2001} simulation.

A second source of error can arise from the scattering of the stars
across the boundary between the Oosterhoff I and metal-rich groups.
However any such scattering would be expected to increase rather than
decrease the metal-rich group scale height.  The calculated scale
height also depends upon the adopted relationship between luminosity
and metallicity.  However, alternative calibrations from the recent
literature give similar results.

Because the metal-rich RR Lyrae stars in Figure \ref{tdmetz} are
distributed relatively close to the Galactic plane compared to the
more metal-poor RRab stars, we already expected these stars to have a
scale height indicative of a disk population.  The fit is, however,
strongly influenced by the stars very close to the plane.  If the RRab stars
within 400 pc of the plane are removed from the sample, leaving
45 RRab stars, and the
calculation is repeated, the resultant scale heights are greater,
about 1.1 kpc but with relatively large uncertainties.  
We can imagine three possible explanations for this:
(1) contamination of the RRab sample close to the plane by some other
type of variable star; (2) small number statistics; or (3) the
presence of a mixture of thin and thick disk stars within the
metal-rich RRab sample.

We believe that the first explanation is unlikely.  The light curves
of the stars within this sample are characteristically those of RRab
stars.  While the light curves of high amplitude $\delta$ Scuti stars
can occasionally resemble those of RRab stars, almost all such stars
have periods shorter than $\sim 0.3$ days.  The second explanation is
more difficult to exclude.  The third explanation is the most 
intriguing of the three -- that the stars comprising our metal-rich
sample are a mixture of stars belonging to an old thin disk and the
thick disk.  When the scale height of the full metal-rich sample is
calculated, the solution in this case would be influenced by the
smaller scale height thin disk component.  Values in the literature
for the scale height of the old thin disk typically run from about 240
to 330 pc (see \citet{Chen:2001} and references therein).  In removing
stars within 400 pc from the sample, we are removing many of the thin
disk stars, leaving a solution dominated by the thick disk component.
Values of the scale height of the old thick disk show some scatter,
from as low as 0.6 to 0.7 \citep{Chen:2001} to about 1.5 kpc
\citep{Gilmore:1983}, but are typically in the neighborhood of 1 kpc.
Our value for the sample of metal-rich RRab stars more than 400 pc
from the plane is comparable to the typical value of the thick disk
scale height.

This is not the first suggestion that the metal-rich RRab stars
contain a mixture of thin and thick disk stars.  \citet{Layden:1995}
obtained a scale height of 0.7 ($+0.5,-0.3$) kpc for RRab stars of
$[Fe/H] > -1$.  However, Layden noted a tendency for the calculated
scale height to decrease with increasing metallicity for stars within
this group.  Although Layden cautioned that the trend was significant
only at the one sigma level, he noted that his scale height was in
between the values often quoted for the old thin and thick disk, and
that this might indicated that his sample contained a mix of these
stars.   One might, however, regard the division of the metal-rich
RRab stars into two populations, thick and old thin disk, to be too
simple.  There might instead just be a trend of decreasing scale
height with increasing metallicity.

More recently, \citet{Maintz:2005} studied the motions of 217
RR Lyrae stars brighter than $V = 12.5$.  They found evidence for a
thick disk component, with a scale height of $1.3 \pm 0.1$ kpc and a
halo component of $4.6 \pm 0.3$ kpc.  In considering their sample of
RR Lyrae stars with $|Z| < 0.5$ kpc, they found some evidence for a
thin disk component with a scale height of $0.38 \pm 0.04$ kpc, though
they caution that they cannot draw a firm distinction between possible
thin and thick disk components.  Some of the stars in the
\citet{Maintz:2005} sample are common to our own, but our approaches
to calculating the scale height are quite different so that the
similarity in our results is significant. From a smaller
sample of RRab stars \citet{Amrose:2001} found a
larger scale height ($1.8 \pm 0.5$ kpc) for the thick disk.  Their
approach, however, was also different from that adopted here, making
no use of metallicity constraints.

\begin{figure}
\figurenum{22}
\includegraphics[angle=-90,scale=0.6]{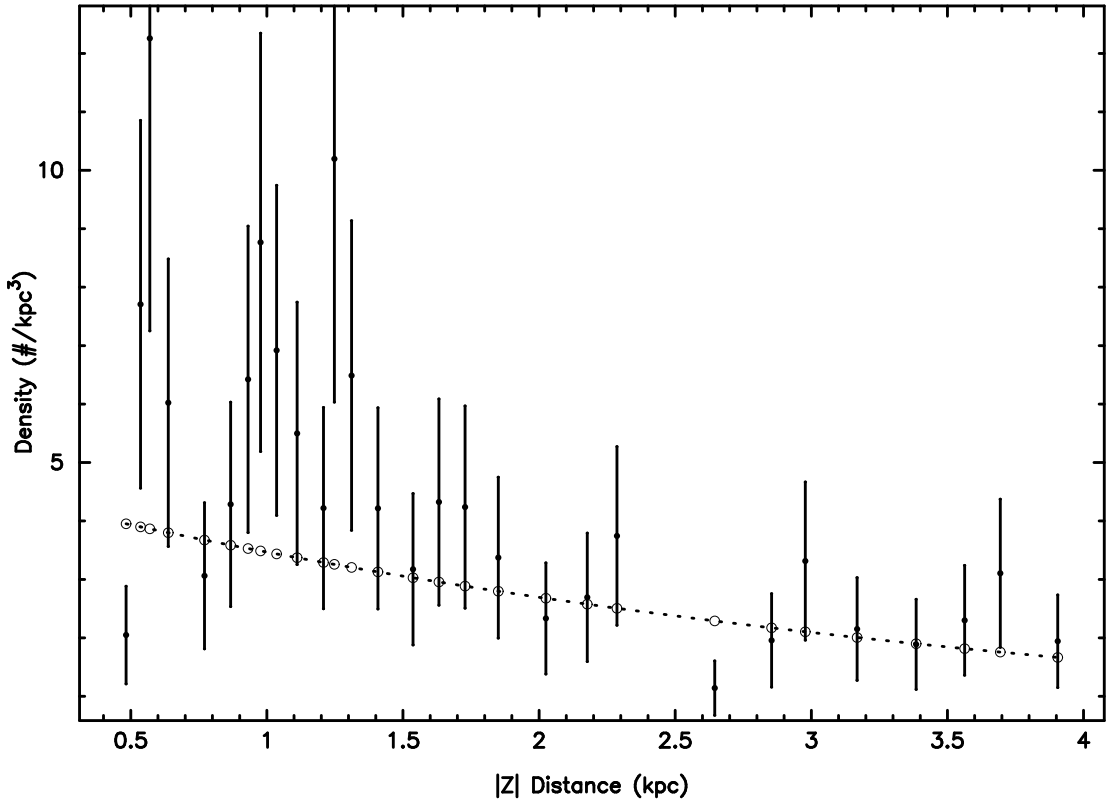}
\caption{The density of Oosterhoff I stars with respect to $|Z|$
  distance is plotted.  The dotted line is the best exponential
  function fit and provides the scale height for this group of stars.}
\label{fig21}
\end{figure}

A vertical scale height may not be the best method for describing the
actual distribution of stars belonging to the Galactic halo.  The
scale height of about 4 kpc found for the Oosterhoff type I sample is
much larger than the scale height of our sample of the metal-rich RRab
stars.  Figure \ref{fig21} shows the scale height fit for the Oosterhoff I
group.  Note that close to the Galactic plane, the data points tend to
fall above the fit to the data as a whole.  This is again, suggestive
of our mixture of thick disk and halo stars within our Oosterhoff type
I sample.  Although our calculated scale height for the ``halo"
group is too uncertain to be very useful here, the \citet{Maintz:2005}
value of 4.6 kpc shows us that our value to be reasonable. 

The presence of RR Lyrae stars is often taken as indicating the
existence of a stellar population with an age of 10 Gyr or more.  If,
in fact, some of the RRab stars in our sample belong to an old thin
disk component, then they may have ages slightly lower than the
canonical 10 Gyr.  \citet{Delpeloso:2005} and \citet{Hansen:2002}
found the age of the old thin disk to be $8.3 \pm 1.8$ Gyr and $7.3
\pm 1.5$ Gyr, respectively.  Thus, it is possible that some of the RR
Lyrae in our sample were formed as recently as 7 or 8 Gyr ago.
However, only a small proportion of the relatively metal-rich red
giants in the old thin disk appear to lose enough mass to produce RR
Lyrae stars \citep{Layden:1995, Taam:1976}.

\subsubsection{$\Delta \log P$ and Metallicity Gradient}

\begin{figure}
\figurenum{23}
\includegraphics[angle=-90,scale=0.6]{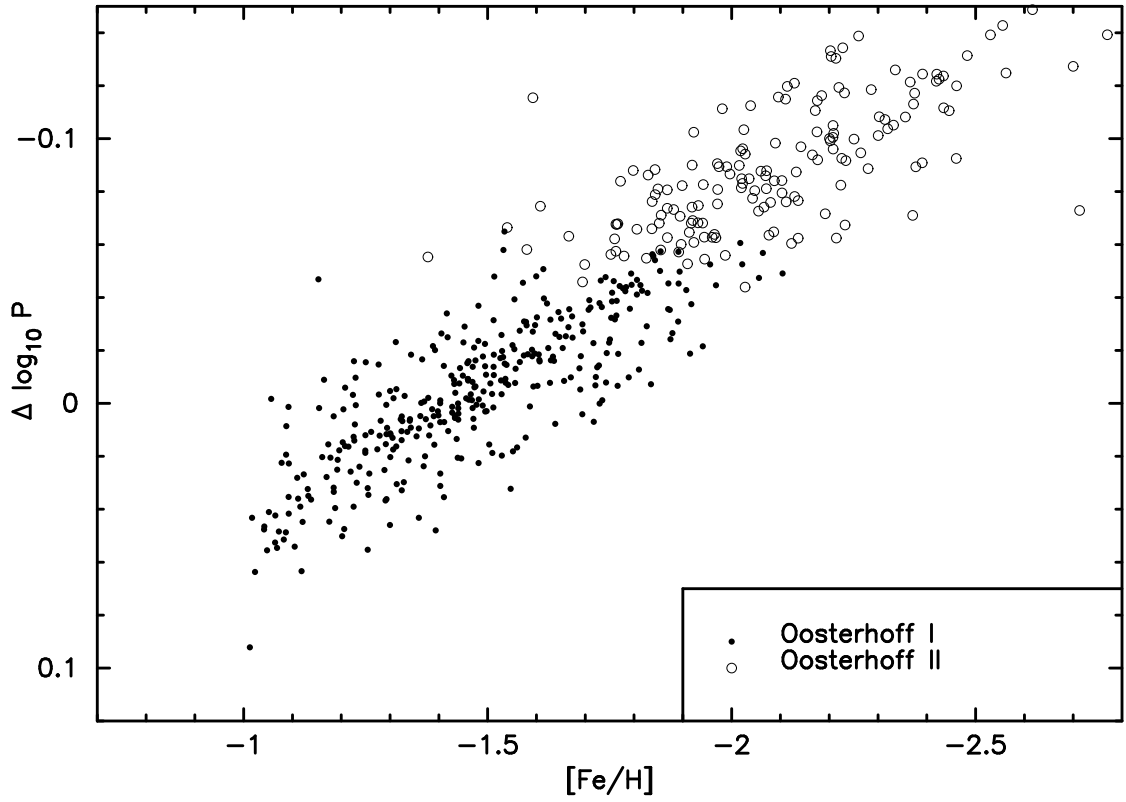}
\caption{This plot is similar to \citet{SKK:1991}'s figure 8a as
  discussed in section 4.4.2.  The
  filled circles are for stars with Oosterhoff I properties, and the
  Oosterhoff II stars are designated by open circles.  Note that
  there is no clear separation between the Oosterhoff groups.}
\label{logpfeh}
\end{figure}

In \citet{SKK:1991} figure 8a, a plot of the period shift,
$\Delta \log P$, with metallicity showed a clear separation of the
field RR Lyraes into the Oosterhoff groups.  Since our
period-amplitude diagram does not clearly show a sharp distinction
between the two Oosterhoff groups, we plotted our sample of RRab stars
in the same manner as \citet{SKK:1991}.  We did not include any of
the metal-rich stars ($[Fe/H] > -1$) that we have identified as our
thick disk sample.  Figure \ref{logpfeh} shows our $\Delta \log
P-[Fe/H]$ plot, with approximately the same axes ranges as in
\citet{SKK:1991}'s figure 8a.  We do not see a gap at $\Delta\log P =
-0.03$, corresponding to the region of the Oosterhoff gap, but rather
a continuous trend from one group to the next Oosterhoff group.
Admittedly, our uncertainties in [Fe/H] are larger on a star to star
basis than those in Suntzeff et al., which will tend to erase any
Oosterhoff gap.

\begin{figure}
\figurenum{24}
\includegraphics[angle=-90,scale=0.6]{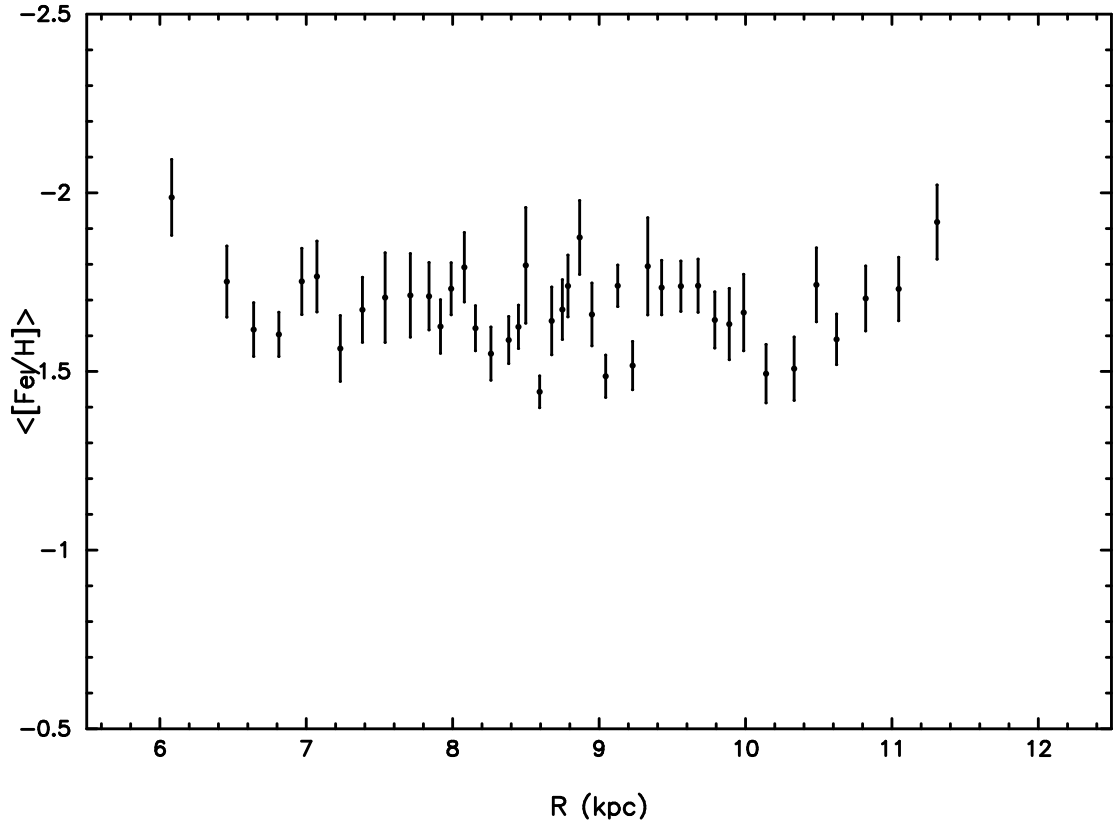}
\caption{Average metallicity distributed with Galactocentric distance,
  R.  With this plot, we investigate the existence of a metallicity
  gradient, which was observed in \citet{SKK:1991}'s data set.  We used the
  same method of binning the stars as described in \citet{SKK:1991}'s
  figure 4.  With our sample of NSVS RRab stars, a metallicity
  gradient could not be clearly discerned.}
\label{skkgrad}
\end{figure}

We also looked for a metallicity gradient as a function of
Galactocentric distance, R, in kpc.  We assume the Galactocentric
distance of the Sun to be 8.5 kpc.  \citet{SKK:1991} reported a
metallicity gradient in their sample of field RR Lyraes, which spanned
a region of $R = 4$ to 10 kpc.  Following their steps, we present our
result in Figure \ref{skkgrad}, however, our sample only covers a
region of $R \sim 6$ to 12 kpc.  In the region that overlaps with
Suntzeff et al's work, we do not see a significant metallicity
gradient.  We note that in Suntzeff et al.'s figure 8a, the greatest 
change in [Fe/H] occurs at roughly $R < 5$ kpc.  
Few stars within our sample are that close to the galactic center.
Although we have tried to exclude obvious disk stars
from Figure \ref{logpfeh}, the \citet{SKK:1991} RRab stars are usually fainter than
the NSVS sample, and may be a more pure halo sample.

\subsection{C-Type RR Lyrae Sample}

Although we have been been reporting our results for the NSVS RRab
stars in this paper, we have also completed a preliminary search for
Bailey type c RR Lyrae (RRc) in the the NSVS database.  This search
for RRc stars was done in conjunction with the search for RRab stars,
but with different selection criteria than those outlined in section
3.1.  However, the 2MASS correlated database from which we obtained
our RRab sample was optimized to find RRab stars and not the lower
amplitude, shorter period RRc stars.  We have found that many of the
GCVS RRc stars were omitted from our preliminary RRc sample.  A
thorough search and analysis of the NSVS RRc stars will be conducted
at a later date.  In this section, we will confine ourselves to a
description of some of the problems encountered with the
identification of RRc stars in the NSVS database. 

We initially obtained 2558 RRc candidates from the 2MASS correlated
NSVS database.  As with the RRab candidates, we removed duplicate
entries but kept the entry with the most observations.  Our selection
criteria for the RRc candidate stars are listed in Table
\ref{selection_rrc}.  Periods were
obtained using the Supersmoother routine and were compared to the
period solutions from the cubic spline method.  After a period was
chosen for each star, we visually inspected these candidates and
removed those stars with phased light curves of an eclipsing binary or
phased light curves of poor quality.  Amplitudes and mean
intensity-weighted magnitudes were calculated for the remaining RRc
candidates using the spline routine used for RRab stars.  Since our
relation for scaling amplitudes was constructed for both RRab and RRc
stars, we applied it to scale the amplitudes of the RRc candidates.
All of the NSVS RRc parameters will be published in a future paper
dealing specifically with these RR Lyrae stars.

We encountered several difficulties in selecting an appropriate RRc
star sample that were more severe than in the case of the corresponding
RRab stars.  The light curve shape typical of RRc stars is not as
distinctive as that of RRab stars.  In particular, when the light
curves are noisy, there can be confusion between the RRc stars, W UMa
type eclipsing binary stars, and short period $\delta$ Scuti/SX
Phoenicis variable stars.  Moreover, there are fewer RRc stars with
excellent NSVS light curves that also have well observed light curves
in the standard Johnson $V$ bandpass.  Thus, we have not yet been able
to compare $\phi_{31}$ values on the NSVS system to standard $V$ band
$\phi_{31}$ values in a satisfactory manner for the RRc stars.

\begin{figure}
\figurenum{25}
\includegraphics[angle=-90,scale=0.6]{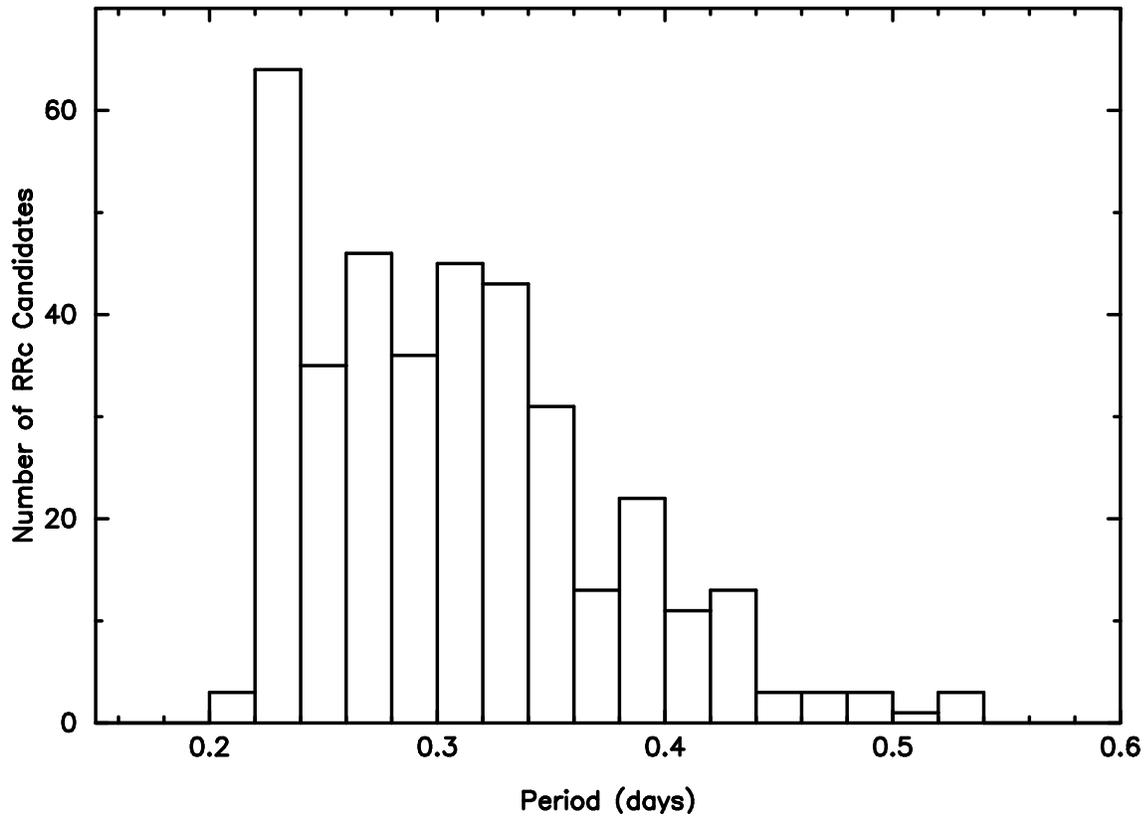}
\caption{Period distribution for 375 NSVS RRc candidate
  stars, as discussed in section 4.6.  Note that in the
  0.22 period bin there appears to be an overabundance of short period
  variables.  This may indicate a contamination of non-RR Lyraes stars
  in our RRc candidate sample.}
\label{rrchist}
\end{figure}

\begin{figure}
\figurenum{26}
\includegraphics[angle=-90,scale=0.6]{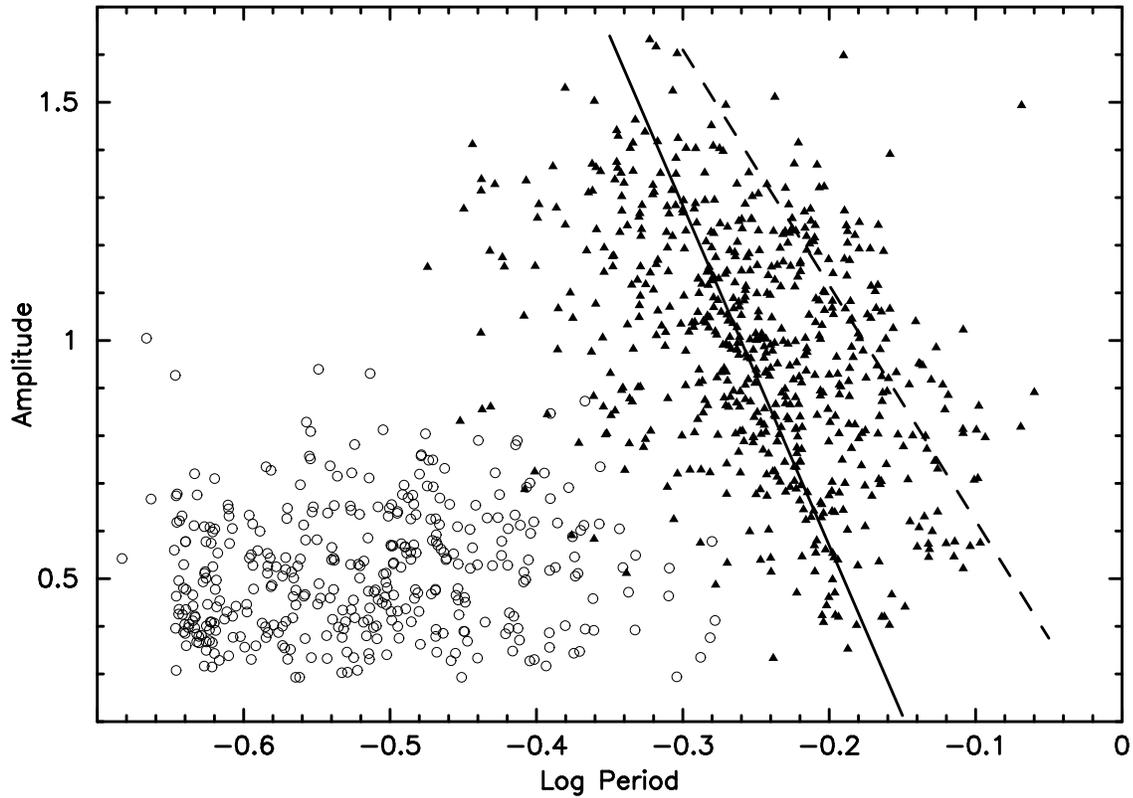}
\caption{The Bailey diagram of the RRab and RRc stars from the NSVS
  survey.  Note the excess of very short period stars in the RRc
  location.  This may be due to a contamination of non-RR Lyrae
  variables such as $\delta$ Scuti stars.  The filled triangles are
  the NSVS RRab stars and the open circles are the NSVS RRc
  candidates.  \citet{Clement:2000}'s trend lines for the Oosterhoff
  dichotomy are overplotted for the NSVS RRab stars.}
\label{pa_rrc}
\end{figure}

Figure \ref{rrchist} shows the period histogram for a sample of 375 RRc
candidates as selected by the criteria in Table \ref{selection_rrc}.  We
also used the selection criteria as described in \citet{Akerlof:2000}
to arrive at this sample.  The sharp increase in the histogram toward
shorter periods is, however, suspicious, indicating a possible
contamination of the RRc stars by longer period, larger amplitude
$\delta$ Scuti type stars.  This possible contamination is also
evident in Figure \ref{pa_rrc}, in which the RRc candidates are
included in the period-amplitude diagram.  The concentration of stars
toward the short period and low amplitude corner of the plot may be
spurious.

So far, attempts to clearly separate RRc stars from the non-RR Lyrae
stars using Fourier decomposition parameters, periods, and amplitudes
have not been entirely successful.  \citet{Wozniak:2004b} have used
machine learning techniques to classify long period variable stars,
and this method may in the future help with the classification of the
short period pulsating variables.  The NSVS RRc variables will
therefore be discussed further in a second paper.

\section{Summary and Future Directions}

Using the Supersmoother algorithm, we derived periods for 1197 RRab
candidates, and obtained the amplitude and mean magnitudes for 608
stars. Photometric metallicities were determined using two methods,
both dependent upon the observed light curves.  The first employs the
Fourier decomposition parameter, $\phi_{31}$ \citep{Jurcsik:96}, whereas the
second relies upon the amplitude \citep{Sandage:2004}. It is apparent
from the period histogram alone, that Oosterhoff type I RR Lyrae stars
dominate the solar neighborhood field population.  However, the
period-amplitude and metallicity information confirm that the
RRab population in the solar neighborhood can be described as a
mixture of metal-rich, Oosterhoff I, and Oosterhoff II stars.

Scale heights calculated for our metal-rich RRab sample ($[Fe/H] > -1$) fully
confirm the supposition that they belong almost entirely to a
disk population.  Some of the RRab stars in the Oosterhoff I group, with
$[Fe/H] < -1$ may also belong to the thick disk population.  There is
some evidence that the metal-rich RRab stars may consist of a thin
disk as well as a thick disk component.  The general conclusion,
however, is that we see a trend of decreasing scale height with
increasing metallicity among the different samples of RRab stars.
Excluding the metal-rich disk sample,  we do not detect a metallicity
gradient with respect to Galactocentric distance.  This may be
slightly at variance with the result of \citet{SKK:1991} who found
their RRab sample to become more metal-rich at lower Galactocentric
distances.  However, our sample is restricted to within a few
kiloparsecs of the sun, and does not extent as near to the galactic
center as that of \citet{SKK:1991}.  We also did not observed a clear
gap between the two Oosterhoff groups when looking at the distribution
of stars with respect to period shift and metallicity. 

We can identify several areas where future studies can expand upon or
clarify issues raised in this paper.  The NSVS database is expected to
provide the first kinematically unbiased and complete RR Lyrae catalog
for stars brighter than $V \sim 15$.  The results presented here are
only for stars found brighter than $V_{ROTSE} = 14$, having more than
40 epochs, and, in many cases, restricted to $|b| > 12^{\circ}$.  The
addition to this study of RR Lyrae candidates fainter than 14th
magnitude in the NSVS database would extend the spatial coverage of
the inner halo RR Lyrae population.  However, for that extension to be
useful, we must have a full understanding of any biases inherent in
the use of fainter RR Lyrae candidates.  As we approach the faint
magnitude limit of the NSVS, we expect to preferentially lose some of
the longer period RRab stars and some RRc stars.  The amplitudes of
these stars are low (anywhere from 0.1 to 0.3 magnitudes), and the
noise in the observations may prevent these stars from being
identified or classified properly.  This bias will affect any
discoveries at these faint magnitudes and possibly influence a
correlation with the Galactic coordinates.  It is also important to
resolve the difficulties in the identification of RRc stars in the
NSVS sample.  Until this is accomplished, an analysis similar to that
done with the RRab sample cannot be completed for the RRc stars.

Many of the RRab stars in our sample lack both radial velocities
and spectroscopic metallicities.  A full kinematic analysis of the thick disk
sample is needed to better establish the properties of the disk component
among both the metal-rich and the Oosterhoff I variables.  With the
kinematic description of the RR Lyraes, we can look for clumps that
may be associated with a tidal stream fragment or subgalactic group
that is accreting into the Galaxy within the solar neighborhood.

Finally, we note that Oaster et al. (2006) have shown that followup photometry 
can be valuable for determining the properties of possible multi-mode
RR Lyrae stars within the NSVS database.  Additional photometry is needed to
determine secondary periods for a number of possible newly identified
Blazhko stars within the RRab sample.  A number of new double mode RR
Lyraes have also been identified \citep{Wils:2006}.

\acknowledgments
K.K. and H.A.S. thank the NSF for support under grant AST-0205813.
K.K. would like to acknowledge support from NSF grant AST-0307778.
H.A.S. also thanks the Center for the Study of Cosmic Evolution.
P.W. was supported by the Oppenheimer Fellowship and the internal LDRD
funding at LANL.  T.A.M. gratefully acknowledges support from NSF
grant AST-040761.  K.K. thanks Tim Beers and Brian Sharpee for many
useful discussions and software support for this project.  The authors
would like to thank the referee, Doug Welch, for his helpful comments
and suggestions to make this a better paper.

This publication makes use of the data from the Northern Sky
Variability Survey created jointly by the Los Alamos National
Laboratory and the University of Michigan.  The NSVS was funded by the
Department of Energy, the National Aeronautics and Space
Administration, and the National Science Foundation.

This publication makes use of data products from the Two Micron All
Sky Survey, which is a joint project of the University of
Massachusetts and the Infrared Processing and Analysis Center, funded
by the National Aeronautics and Space Administration and the National
Science Foundation.

\bibliography{apj-jour,references}

\clearpage

\clearpage

\clearpage

\begin{deluxetable}{cc}
\tablewidth{0pc}
\tablecaption{RRab candidate criteria for selection from the NSVS
survey.}
\tablenum{1}
\tablehead{\colhead{Parameter} & \colhead{Range of Values}}
\startdata
Period & 0.30 to 0.95 days\\
Amplitude & 0.20 to 1.5 magnitudes \\
Ratio & $> -0.625(Period) + 0.719$\\
$J-H$ & -0.1 to 0.5 \\
$H-K$ & -0.1 to 0.25\\
\enddata
\label{selection_rrab}
\end{deluxetable}

\clearpage
\begin{deluxetable}{ccccc}
\tablecaption{Undetermined periods for 6 RRab stars.}
\tablenum{2}
\tablewidth{0pc}
\tablehead{\colhead{NSVS id} & \colhead{RA(2000)} &
  \colhead{DEC(2000)} & \colhead{Period 1} & \colhead{Period 2} \\ 
\colhead{} & \colhead{} & \colhead{} & \colhead{(days)} & \colhead{(days)} } 
\startdata
11570713 & 318.6590 & 7.6242  & 0.46169 & 0.31593 \\
11647538 & 322.7000 & 17.6452 & 0.37979 & 0.61282 \\
15961396 & 187.7340 & -14.0363 & 0.54893 & 0.35358 \\
16061651 & 208.2150 & -13.8435 & 0.54246 & 0.35138 \\
17104793\tablenotemark{a} & 303.9150 & -12.4860 & 0.57206 & 0.36360 \\
17575093 & 45.3470 & -28.7829 & 0.59718 & 0.37357 
\enddata
\tablenotetext{a}{The All Sky
  Automated Survey (ASAS) \citep{Pojmanski:2002} variable star catalog does
  have a listing for this star, and the ASAS listed period for
  this star is 0.57199 days.}

\label{unknown_p}
\end{deluxetable}

\clearpage
\begin{deluxetable}{ccccccc}
\tablenum{3}
\tablewidth{0pc}
\tablecaption{Photometric Metallicity of NSVS RRab Stars.}
\tablehead{\colhead{NSVS ID} & \colhead{$[Fe/H]_{AP}$} & \colhead{$\sigma_{AP}$} & \colhead{$[Fe/H]_{JK}$} & \colhead{$\sigma_{JK}$} & \colhead{[Fe/H]} & \colhead{$\sigma_{[Fe/H]}$} }

\startdata
47457 & -1.47 & 0.32 & -1.84 &  0.19 & -1.73 &  0.16 \\
95599 & -2.17 & 0.32 & -2.32 &  0.19 & -2.28 &  0.16 \\
219255 & -1.16 & 0.32 &  ...  &   ... & -1.16 &  0.32 \\
253439 & -1.71 & 0.32 &  ...  &   ... & -1.71 &  0.32 \\
258744 & -0.49 & 0.32 & -0.06 &  0.29 & -0.27 &  0.21 \\
272580 & -1.20 & 0.32 & -1.38 &  0.23 & -1.31 &  0.18 

\enddata
\label{rrab_feh}
\tablecomments{A complete version of this table is available
  in the electronic version of the Journal.  This table is only a
  sample and should be used as an aide for the full table.}
\end{deluxetable}

\clearpage
\begin{deluxetable}{cccccc}
\tablecaption{NSVS RRab Properties.}
\tablenum{4}
\tablewidth{0pc}

\tablehead{\colhead{NSVS ID} & \colhead{RA (2000)} & \colhead{DEC (2000)} & \colhead{Period} & \colhead{Amp ($V_{ROTSE}$)} & \colhead{$\langle V \rangle$} \\ 
\colhead{} & \colhead{} & \colhead{} & \colhead{(days)} & \colhead{} & \colhead{} } 
\startdata
47457 & 341.080994 & 83.949997 & 0.52608 & 1.07 & 12.47 \\
94165 & 154.889999 & 83.603500 & 0.59265 & 1.40 & 14.45 \\
95599 & 125.681000 & 86.081200 & 0.65198 & 1.04 & 12.57 \\
219255 & 13.649000 & 74.527901 & 0.64964 & 0.35 & 10.62 \\
253439 & 7.187000 & 80.033997 & 0.61916 & 0.84 & 13.88 
\enddata
\label{rrab_properties}
\tablecomments{A complete version of this table is available
  in the electronic version of the Journal.  This table is only a
  sample and should be used as an aide for the full table.}
\end{deluxetable}

\clearpage

\begin{deluxetable}{ccc}
\tablecaption{Summary of the number ratio of stars in Oosterhoff
  groups.}
\tablenum{5}
\tablewidth{0pc}
\tablehead{\colhead{Number Ratio} & \colhead{$|Z| < 2.0$ kpc} & \colhead{$2.5 < |Z| < 3.5$ kpc} \\ 
\colhead{} & \colhead{} & \colhead{} } 
\startdata
Oo II/Oo I & $0.37 \pm 0.05$ & $0.57 \pm 0.12$ \\
Oo II/Oo I (metal-poor) & $0.47 \pm 0.06$ & $0.67 \pm 0.15$ \\
Oo II/Oo I (metal-rich) & $1.5 \pm 0.3$ & $3.9 \pm 1.4$ 
\enddata
\label{numratio}
\end{deluxetable}

\clearpage

\begin{deluxetable}{cccccccc}
\tablecaption{Number distribution of RRab stars into Oosterhoff and metal-rich groups.}
\tablenum{6}
\tablewidth{0pc}
\tablehead{\colhead{$|Z|$ Region} & \colhead{Metal-Rich} & \colhead{Oo I} & \colhead{Oo II} & \colhead{Metal-Rich} & \colhead{Oo I} & \colhead{Oo II} \\ 
\colhead{(kpc)} & \colhead{} & \colhead{} & \colhead{} & \colhead{($|b| > 12^{\circ}$)} & \colhead{($|b| > 12^{\circ}$)} & \colhead{($|b| > 12^{\circ}$)} } 

\startdata
0.0-0.25 & 11 & 13 & 1 & 3 & 2 & 0 &  \\
0.25-0.50 & 19 & 19 & 9 & 13 & 13 & 7 \\
0.50-0.75 & 12 & 27 & 8 & 12 & 27 & 7 \\
0.75-1.0 & 11 & 35 & 8 & 11 & 35 & 8 \\
1.0-1.25 & 4 & 43 & 14 & 4 & 43 & 14 \\
1.25-1.5 & 8 & 39 & 8 & 8 & 39 & 8 \\
1.5-1.75 & 1 & 24 & 12 & 1 & 24 & 12 \\
1.75-2.0 & 2 & 27 & 22 & 2 & 27 & 22 \\
2.0-2.25 & 1 & 25 & 17 & 1 & 25 & 17 \\
2.25-2.5 & 0 & 19 & 17 & 0 & 19 & 17 \\
2.5-2.75 & 0 & 16 & 8 & 0 & 16 & 8 \\
2.75-3.0 & 0 & 20 & 9 & 0 & 20 & 9 \\
3.0-3.25 & 0 & 16 & 11 & 0 & 16 & 11 \\
3.25-3.5 & 0 & 9 & 7 & 0 & 9 & 7 \\
3.5-3.75 & 0 & 11 & 6 & 0 & 11 & 6 \\
3.75-4.0 & 0 & 4 & 6 & 0 & 4 & 6 \\
4.0-4.25 & 0 & 1 & 3 & 0 & 1 & 3 \\
4.25-4.5 & 0 & 1 & 0 & 0 & 1 & 0 \\
4.5-4.75 & 0 & 0 & 0 & 0 & 0 & 0 \\
4.75-5.0 & 0 & 0 & 0 & 0 & 0 & 0
\enddata
\label{oo_numbers}
\end{deluxetable}


\begin{deluxetable}{cccccc}
\tablecaption{Number of RRab stars in the Oosterhoff I subgroups.}
\tablenum{7}
\tablewidth{0pc}
\tablehead{\colhead{$|Z|$ Zone} & \colhead{Oo I-poor} & \colhead{Oo I-rich} & \colhead{Oo I-poor } & \colhead{Oo I-rich} \\ 
\colhead{(kpc)} & \colhead{} & \colhead{} & \colhead{$|b| > 12^{\circ}$} & \colhead{$|b| > 12^{\circ}$} }

\startdata
0.0-0.25 & 12 & 1 & 2 & 0 \\

0.25-0.5 & 11 & 8 & 8 & 5 \\

0.5-0.75 & 23 & 4 & 23 & 4 \\

0.75-1.0 & 23 & 12 & 23 & 12 \\

1.0-1.25 & 31 & 12 & 31 & 12 \\

1.25-1.5 & 29 & 10 & 29 & 10 \\

1.5-1.75 & 19 & 5  & 19 & 5 \\

1.75-2.0 & 25 & 2 & 25 & 2  \\

2.0-2.25 & 24 & 1  & 24 & 1 \\

2.25-2.5 & 17 & 2  & 17 & 2 \\

2.5-2.75 & 13 & 3  & 13 & 3 \\

2.75-3.0 & 17 & 3  & 17 & 3 \\

3.0-3.25 & 14 & 2  & 14 & 2 \\

3.25-3.5 & 8 & 1  & 8 & 1 \\

3.5-3.75 & 10 & 1 & 10 & 1  \\

3.75-4.0 & 4 & 0  & 4 & 0 \\

4.0-4.25 & 1 & 0 & 1 & 0  \\

4.25-4.5 & 0 & 0 & 0 & 0  \\

4.5-4.75 & 0 & 0  & 0 & 0 \\

4.75-5.0 & 0 & 0  & 0 & 0
\enddata
\label{ooI_num}
\end{deluxetable}


\begin{deluxetable}{ccccc}
\tablecaption{Scale heights of different Galactic groups of RRab stars. }
\tablenum{8}
\tablewidth{0pc}
\tablehead{\colhead{Group} & \colhead{No. of stars} & \colhead{h(Z) [6 stars/bin]} & \colhead{h(Z) [10 stars/bin]} \\ 
\colhead{} & \colhead{} & \colhead{(kpc)} & \colhead{(kpc)} } 

\startdata
Metal-rich & 69 & $0.65 \pm 0.17$ & $0.68 \pm 0.18$  \\
Metal-rich ($|Z| > 0.4$ kpc) & 45 & $1.04 \pm 0.48$ & $1.22 \pm 0.65$ \\
Metal-rich (detector efficiency) & 69 & $0.66 \pm 0.16$ & $0.67 \pm 0.17$ \\
Metal-rich ($|b| > 12^{\circ}$ only) & 55 & $0.77 \pm 0.24$ & $0.90 \pm 0.47$ \\
Oosterhoff I & 330 & $4.0 \pm 1.3$ & $3.9 \pm 1.3$ \\
Halo & 428 & $6.9 \pm 3.3$ & $5.5 \pm 2.1$
\enddata
\label{sclhgt}
\end{deluxetable}

\clearpage
\begin{deluxetable}{ccc}
\tablewidth{0pc}
\tablecaption{RRc candidate criteria for selection from the NSVS
survey.}
\tablenum{9}
\tablehead{\colhead{Parameter} & \colhead{Range in Triangular
region} & \colhead{Range in Rectangular region}}
\startdata
Period & 0.225 to 0.3 days & 0.3 to 0.55 days\\
Amplitude & 0.2 to 0.7 magnitudes &  0.2 to 0.7\\
Ratio & 0.2 to 0.8 & $> 0.2$ and $< (-0.625(Period) + 0.719)$\\
$J-H$ & -0.1 to 0.5 &  -0.1 to 0.5\\
$H-K$ & -0.1 to 0.25 & -0.1 to 0.25\\
\enddata
\label{selection_rrc}
\end{deluxetable}

\end{document}